\documentclass[article, nojss]{jss}
\usepackage{orcidlink,thumbpdf,lmodern}
\usepackage{framed}

\usepackage{amsmath, amssymb}
\usepackage{dsfont}
\usepackage{enumitem}
\usepackage{tikz}
\usetikzlibrary{arrows, positioning, shapes}
\usepackage{placeins}

\newcommand{\class}[1]{`\code{#1}'}

\newcommand{\bs}{\boldsymbol}
\newcommand{\N}{\mathds{N}}
\newcommand{\R}{\mathds{R}}
\DeclareMathOperator*{\argmin}{argmin}

\makeatletter
\newcommand{\Rg}{\mathcal{R}} % Main symbol: R (in calligraphic font)
\newcommand{\Rf}{\@ifnextchar*{\Rf@opt}{\Rf@noopt}}
\newcommand{\Rf@opt}[1]{\@ifnextchar\bgroup{\Rf@opt@child}{\Rg_{F}}}
\newcommand{\Rf@noopt}{\@ifnextchar\bgroup{\Rf@child}{\Rg_{F^m}}}
\newcommand{\Rf@opt@child}[1]{\@ifnextchar\bgroup{\Rf@opt@index{#1}}{\Rg_{F_{#1}}}}
\newcommand{\Rf@opt@index}[2]{\Rg_{{F_{#1}}#2}}
\newcommand{\Rf@child}[1]{\@ifnextchar\bgroup{\Rf@index{#1}}{\Rg_{F_{#1}^m}}}
\newcommand{\Rf@index}[2]{\Rg_{{F_{#1}}#2^m}}
\makeatother

\author{
Taiane S. Prass~\orcidlink{0000-0003-3136-909X}\\
PPGEst - UFRGS
\And
Alisson S. Neimaier~\orcidlink{0000-0002-7524-0776}\\
PPGEst - UFRGS
\And
Guilherme Pumi~\orcidlink{0000-0002-6256-3170}\\
PPGEst - UFRGS
}

\Plainauthor{Taiane Schaedler Prass, Alisson Silva Neimaier, Guilherme Pumi}

\title{PRTree: An \proglang{R} Package for Probabilistic Regression Trees with Built-in Missing Data Handling}

\Plaintitle{PRTree: An R Package for Probabilistic Regression Trees with Built-in Missing Data Handling}

\Shorttitle{\pkg{PRTree}: Probabilistic Regression Trees in \proglang{R}}

\Abstract{
\pkg{PRTree} is an \proglang{R} package for fitting Probabilistic Regression Trees (PRTrees), a class of regression trees that replaces deterministic splits with probabilistic associations to produce smooth prediction functions. The package implements both the original methodology and its recent extension for handling missing predictor values, allowing model fitting and prediction directly from incomplete datasets without prior imputation. It provides a unified framework for model fitting, prediction, visualization, smoothing-parameter selection, cross-validation, and model diagnostics through a standard \proglang{R} interface. Computationally intensive routines are implemented in \proglang{FORTRAN}, while the high-level interface follows the usual \proglang{R} workflow based on S3 classes and generic methods. This paper reviews the underlying methodology, describes the software architecture and main package components, and illustrates their use through reproducible examples.
}

\Keywords{probabilistic regression trees, missing data, nonparametric regression, statistical software, \proglang{R}}

\Plainkeywords{probabilistic regression trees, missing data, nonparametric regression, statistical software, R}

\Address{
  Taiane Schaedler Prass; Guilherme Pumi\\
  Department of Statistics\\
  Graduate Program in Statistics (PPGEst)\\
  Mathematics and Statistics Institute (IME)\\
  Universidade Federal do Rio Grande do Sul (UFRGS)\\
  9500, Bento Gonçalves Avenue, Porto Alegre - RS, Brazil\\
  E-mail: \email{taiane.prass@ufrgs.br (Prass); guilherme.pumi@ufrgs.br (Pumi)}\\
  URL: \url{https://professor.ufrgs.br/tsprass}
}

\begin{document}

%% -- Introduction -------------------------------------------------------------

%% - But should typically have some discussion of both _software_ and _methods_.
%% - Use \proglang{}, \pkg{}, and \code{} markup throughout the manuscript.
%% - If such markup is in (sub)section titles, a plain text version has to be
%%   added as well.
%% - All software mentioned should be properly \cite-d.
%% - All abbreviations should be introduced.
%% - Unless the expansions of abbreviations are proper names (like "Journal
%%   of Statistical Software" above) they should be in sentence case (like
%%   "generalized linear models" below).

\section{Introduction}

Regression trees are among the most widely used nonparametric methods for supervised learning because they naturally capture nonlinear relationships and complex interactions while producing interpretable predictive models. Since the introduction of Classification and Regression Trees (CART) by \citet{breiman84}, tree-based methods have become a standard tool in statistics and machine learning, with mature implementations available in the \proglang{R} ecosystem, most notably through the \pkg{rpart} package \citep{rpart}. Classical regression trees, however, partition the predictor space through deterministic recursive binary splits, producing piecewise-constant prediction functions that may inadequately represent smooth underlying regression functions.

Several extensions have been proposed to overcome this limitation by replacing hard decisions with smooth probabilistic mechanisms. Smooth Transition Regression Trees (STR-Trees; \citealp{medeiros2008}) replace the hard splits of CART by logistic transition functions while preserving axis-aligned splits based on a single predictor variable at each internal node. An implementation of the STR-Trees for \proglang{R} is available from GitHub (\url{https://github.com/gabrielrvsc/BooST}). Soft Trees \citep{irsoy2012} determine routing probabilities through sigmoid functions of linear combinations of the predictor variables, allowing oblique probabilistic splits. A reference C++ implementation is available at \url{https://github.com/oir/soft-tree}. Probabilistic Regression Trees (PRTrees), introduced by \citet{alkhoury2020}, adopt a different perspective by directly defining probability measures over regions of the predictor space through smoothing densities. The reference Python implementation is available at \url{https://gitlab.com/sami.courie/pr-tree}.

While differing in their probabilistic formulations, these methods share the common goal of replacing the discontinuous predictions of classical regression trees with smooth prediction functions. However, they all assume that the predictor vector is fully observed and therefore provide no native mechanism for handling missing predictor values during model fitting. In the companion methodological paper \citet{PRTree2026_theory}, the original PRTree algorithm is adapted to accommodate missing predictor values by introducing alternative definitions of the region-association probabilities for partially observed predictor vectors. These adaptations preserve the probabilistic structure of PRTrees while allowing model fitting and prediction directly from incomplete data without requiring external imputation.

To make these developments readily available to researchers and practitioners, this paper presents \pkg{PRTree}, the first public \proglang{R} implementation of Probabilistic Regression Trees. The package implements both the original methodology of \citet{alkhoury2020} and the missing-data extensions proposed by \citet{PRTree2026_theory} within a unified software framework. It combines a high-level \proglang{R} interface with computational kernels implemented in FORTRAN and C, providing an integrated workflow for model fitting, prediction, visualization, cross-validation, smoothing-parameter selection, and reproducible empirical evaluation. The \pkg{PRTree} package is the first publicly available software implementing the proposed missing-data extensions for Probabilistic Regression Trees and, to the best of our knowledge, the first implementatio of the PRTree algorithm for \proglang{R}.

The remainder of this paper is organized as follows. Section~\ref{sec:prtree_method} briefly reviews the Probabilistic Regression Tree methodology and the missing-data strategies implemented in \pkg{PRTree}. Section~\ref{sec:package} describes the main components of the package and its implementation. Section~\ref{sec:examples} illustrates the principal functionalities of \pkg{PRTree} through reproducible examples. Finally, Section~\ref{sec:summary} concludes with a discussion of the package and directions for future development.

\section{Probabilistic Regression Tree Model}\label{sec:prtree_method}

This section briefly reviews the basic concepts underlying the methodology implemented in \pkg{PRTree}. These concepts are sufficient to understand the package and the examples presented throughout the paper. Complete methodological details of the original PRTree algorithm and its missing-data extension can be found in \citet{alkhoury2020} and \citet{PRTree2026_theory}, respectively.

\subsection{PRTree Methodology}

Let $(\bs{X},Y)$ be a random vector taking values in $\R^p\times\R$, where $Y$ denotes the response variable and $\bs{X}=(X_1,\cdots,X_p)$ the predictor vector. The regression function is defined by $m(\bs{x})=\E(Y|\bs{X}=\bs{x})$. A PRTree recursively partitions the predictor space into $M$ disjoint rectangular regions $\Rg_1,\cdots,\Rg_M$. Unlike classical regression trees, an observation is not deterministically assigned to a single terminal region. Instead, for every measurable region $\Rg\subset\R^p$ and predictor value $\bs{x}$, the method defines the probability of association
\begin{equation}\label{eq:Psi}
\Prob(\Rg|\bs{X}=\bs{x})=\biggl[\prod_{k=1}^{p}\sigma_k\biggr]^{-1}\int_{\Rg}\phi\biggl(\frac{v_1-x_1}{\sigma_1},\cdots,\frac{v_p-x_p}{\sigma_p}\biggr)\,d\bs{v},
\end{equation}
where $\phi$ is a probability density function and $\bs{\sigma}=(\sigma_1,\cdots,\sigma_p)$ is a vector of smoothing parameters. Following \citet{alkhoury2020}, we denote the association probability by $\Psi(\bs{x};\Rg,\bs{\sigma})=\Prob(\Rg|\bs{X}=\bs{x})$.
 The package currently supports Gaussian, Log-normal, Student's $t$, and Gamma association densities.

The PRTree regression estimator is
\[
\hat m(\bs{x})=\sum_{m=1}^{M}\gamma_m\Psi(\bs{x};\Rg_m,\bs{\sigma}),
\]
where $\gamma_m\in\R$ is the coefficient associated with terminal region $\Rg_m$. Since $\Rg_1,\cdots,\Rg_M$ form a partition of the predictor space, $\sum_{m=1}^{M}\Psi(\bs{x};\Rg_m,\bs{\sigma})=1$ for every $\bs{x}\in\R^p$. Consequently, $\hat m(\bs{x})$ is a convex combination of the coefficients $\gamma_1,\cdots,\gamma_M$, with weights given by the association probabilities.

Model fitting proceeds by recursively partitioning the predictor space. At each iteration, candidate splits are evaluated according to the empirical loss, the terminal-node weights are updated, and tree growth continues until a stopping criterion is satisfied. The estimation of the model parameters is described in Section~\ref{sec:parameter}.

\subsection{Handling Missing Data in PRTrees}\label{sec:missing_handling}

The missing-data methodology implemented in \pkg{PRTree} follows the probabilistic framework proposed in \citet{PRTree2026_theory}. When all predictor values are observed, the association probability $\Psi(\bs{X}_i;\Rg_m,\bs{\sigma})$ is evaluated directly. However, when one or more predictor values are missing, this quantity is no longer directly defined because the complete predictor vector is unavailable. To overcome this limitation, \citet{PRTree2026_theory} introduces three alternative extensions of the association probability for partially observed predictor vectors. These extensions differ only in how $\Psi$ is evaluated; the remaining tree construction algorithm is unchanged. All three approaches are implemented in the package through the \code{fill_type} argument. The adapted function $\Psi^\ast$ (and, consequently, the corresponding probability measure $P^\ast$) is defined recursively as follows.

To fix notation, let $S \subseteq \{1,\cdots,p\}$ denote an arbitrary subset of indices. For any $\bs X \in \R^p$, denote by $\bs X_{|S}$ the subvector containing only the coordinates indexed by $S$. For any rectangular region $\Rg_m=\prod_{j=1}^{p}\Rg_{mj}\subset\R^p$, denote by $\Rg_{m|S}=\prod_{j\in S}\Rg_{mj}$ its projection onto these coordinates, and for any vector $\bs{\sigma}\in\R_+^p$, let $\bs{\sigma}_{|S}$ be its restriction to the corresponding components.

By a slight abuse of notation, the same symbol $\Psi$ is used to denote both the original association probability in Equation~\eqref{eq:Psi} and its restriction to lower-dimensional predictor spaces. Thus, whenever restricted arguments such as $\bs X_{|S}$, $\Rg_{|S}$, and $\bs{\sigma}_{|S}$ are used, $\Psi$ is understood to be defined by Equation~\eqref{eq:Psi} after replacing the predictor vector, the region, and the smoothing vector by their corresponding restrictions.

A notational convention used throughout the paper requires explicit mention: $I(A)$ denotes the indicator function of event $A$ and, by convention, is always evaluated first in any expression where it appears. This convention ensures unambiguous interpretation of formulas involving products of indicators with other mathematical objects.

\paragraph{Adapted association probability.}

Let $\Psi$ be the function defined by Equation~\eqref{eq:Psi}. Given $\bs{X}\in\R^p$ and any region $\Rg\subseteq\R^p$:

\noindent\textbf{(a) Base case:} if $\Rg=\R^p$ (root node), then
\[
\Psi^\ast(\bs{X};\Rg,\bs{\sigma})=1.
\]

\noindent\textbf{(b) Recursive step:} if $\Rg\in\{\Rf*{L},\Rf*{R}\}$ is a child region resulting from the split of a parent region $\Rf*$, first define the set of indices $S(\bs{X},\Rg)$ as follows:
\[
S(\bs{X},\Rg):=\bigl\{j\in\{1,\cdots,p\}:\Rf*{L}{j}\subsetneq\R\text{ and }X_j\text{ is non-missing}\bigr\},
\]
i.e., the indices $j$ such that the corresponding variable has been used in a previous split and $X_j$ is non-missing. Next, define the proxy association function $H$ as
\begin{equation*}
H(\bs{X};\Rg,\bs{\sigma})=
\begin{cases}
\Psi(\bs{X};\Rg,\bs{\sigma}), & \text{if }\bs{X}\text{ is fully observed},\\[4pt]
1, & \text{if \code{fill\_type = 0} or }S(\bs{X},\Rg)=\emptyset,\\[4pt]
I(\bs{X}_{|S(\bs{X},\Rg)}\in\Rg_{|S(\bs{X},\Rg)}), & \text{if \code{fill\_type = 1} and }S(\bs{X},\Rg)\neq\emptyset,\\[4pt]
\Psi(\bs{X}_{|S(\bs{X},\Rg)};\Rg_{|S(\bs{X},\Rg)},\bs{\sigma}_{|S(\bs{X},\Rg)}), & \text{if \code{fill\_type = 2} and }S(\bs{X},\Rg)\neq\emptyset.
\end{cases}
\end{equation*}
Then $\Psi^\ast$ is given recursively by setting $p:=\Psi^\ast(\bs{X};\Rf*,\bs{\sigma})$ and computing
\[
\Psi^\ast(\bs{X};\Rf*{L},\bs{\sigma})=w_1p,\quad
\Psi^\ast(\bs{X};\Rf*{R},\bs{\sigma})=w_2p,
\]
where $w_1=w_2=0$ if $p=0$, and
\[
w_1=\frac{H(\bs{X};\Rf*{L},\bs{\sigma})}{H(\bs{X};\Rf*{L},\bs{\sigma})+H(\bs{X};\Rf*{R},\bs{\sigma})},\quad
w_2=\frac{H(\bs{X};\Rf*{R},\bs{\sigma})}{H(\bs{X};\Rf*{L},\bs{\sigma})+H(\bs{X};\Rf*{R},\bs{\sigma})},\quad
\text{if }p>0.
\]

Figure~\ref{fig:missing_tree} illustrates the three missing-data strategies for a two-dimensional predictor space when the second coordinate is missing. Panel (a) shows a PRTree and panel (b) the corresponding partition of the predictor space. The gray dot represents the true observation $(x_1,x_2)$, while the dashed vertical line represents all possible values of the unobserved coordinate that are compatible with the observed information $(x_1,?)$. Although the terminal regions are known, the original association probability $\Psi(\bs{x};\Rg,\bs{\sigma})$ cannot be evaluated because its definition depends on the complete predictor vector. For illustration, let $p=\Prob(\Rg_2\cup\Rg_3|\bs{x})$ denote the probability associated with the projected region. Panel (c) shows how the three missing-data strategies redistribute this probability between $\Rg_2$ and $\Rg_3$. For \code{fill_type = 0}, the weights are fixed at $w_1=w_2=1/2$. For \code{fill_type = 1}, they are given by $w_1=1$ and $w_2=0$. For \code{fill_type = 2}, the weights are proportional to the probability mass of the association density over the projected regions, represented by the shaded areas under the curve. The resulting association probabilities are then $\Prob(\Rg_2|\bs{x})=w_1p$ and $\Prob(\Rg_3|\bs{x})=w_2p$. All three strategies reduce to the original PRTree formulation when no predictor values are missing. The default implementation of \pkg{PRTree} uses \code{fill_type = 2}.

\begin{figure}[!ht]
\centering
\begin{tikzpicture}[>=latex,font=\small]

%%%%%%%%%%%%%%%%%%%%%%%%%%%%%%%%%%%%%%%%%%%%%%%%%%%%%%%%%%%%
%% (a) Tree
%%%%%%%%%%%%%%%%%%%%%%%%%%%%%%%%%%%%%%%%%%%%%%%%%%%%%%%%%%%%

\begin{scope}
\node at (2.2,5.2){\textbf{(a) Tree structure}};
\node[circle,draw,minimum size=5mm] (r) at (2,4.5) {};
\node[circle,draw,minimum size=5mm] (l) at (1.0,3.5) {};
\node[circle,draw,minimum size=5mm] (rr) at (3,3.5) {};
\node[circle,draw,minimum size=5mm] (b) at (2,2.5) {};
\node[draw,rectangle] (R1) at (0.8,2.0) {$\Rg_1$};
\node[draw,rectangle] (R2) at (1.9,1.2) {$\Rg_2$};
\node[draw,rectangle] (R3) at (3.4,1.2) {$\Rg_3$};
\node[draw,rectangle] (R4) at (4.0,2.5) {$\Rg_4$};
\draw (r)--(l);
\draw (r)--(rr);
\draw (l)--(R1);
\draw (rr)--(b);
\draw (rr)--(R4);
\draw (b)--(R2);
\draw (b)--(R3);
\end{scope}

%%%%%%%%%%%%%%%%%%%%%%%%%%%%%%%%%%%%%%%%%%%%%%%%%%%%%%%%%%%%
%% (b) Partition
%%%%%%%%%%%%%%%%%%%%%%%%%%%%%%%%%%%%%%%%%%%%%%%%%%%%%%%%%%%%
\begin{scope}[xshift=5.3cm]
\node at (2,5.2){\textbf{(b) Partition}};
\draw[->] (0,0)--(4.6,0) node[right]{$x_1$};
\draw[->] (0,0)--(0,4.6) node[above]{$x_2$};
\draw (0,0) rectangle (4,4);
\draw (1.5,0)--(1.5,4);
\draw (1.5,2.2)--(4,2.2);
\draw (3,0)--(3,2.2);
\node at (0.75,2){$\Rg_1$};
\node at (2.75,3.1){$\Rg_4$};
\node at (2.35,1){$\Rg_2$};
\node at (3.5,1){$\Rg_3$};
% true (unobserved) location
\fill[gray!60] (2,1.45) circle (2pt);
% observed x1
\draw[dashed,thick] (2,0)--(2,4);
\node[above] at (2,4){$(x_1,?)$};
\node[above] at (2,4){$(x_1,?)$};
\draw[dashed,thick] (2,0)--(2,4);
\end{scope}

%%%%%%%%%%%%%%%%%%%%%%%%%%%%%%%%%%%%%%%%%%%%%%%%%%%%%%%%%%%%
%% (c)
%%%%%%%%%%%%%%%%%%%%%%%%%%%%%%%%%%%%%%%%%%%%%%%%%%%%%%%%%%%%
\begin{scope}[xshift=10.5cm]
\node at (2.2,5.2){\textbf{(c) Redistribution of $p$}};
\node[align=left,font=\small,anchor=west]  at (0.5, 4.5) {
$p=P(\Rg_2\cup\Rg_3 | \bs{x})$};

%%%%%%%%%%%%%%%%%%%%%%%%%%%%%%%%%%%%
%% fill 0
%%%%%%%%%%%%%%%%%%%%%%%%%%%%%%%%%%%%
\node[anchor=west] at (0,3.8){fill = 0};
\draw[line width=5pt] (1.5,3.8)--(4,3.8);
\draw[dashed] (3,3.4)--(3,3.8);
\node at (2.2,3.25){$w_1 = \frac12$};
\node at (3.8,3.25){$w_2 = \frac12$};

%%%%%%%%%%%%%%%%%%%%%%%%%%%%%%%%%%%%
%% fill 1
%%%%%%%%%%%%%%%%%%%%%%%%%%%%%%%%%%%%
\node[anchor=west] at (0,2.5){fill = 1};
\draw[line width=5pt] (1.5,2.5)--(4,2.5);
\fill[gray!60] (1.5,2.35) rectangle (3,2.65);
\draw[dashed] (3,2.1)--(3,2.5);
\node at (2.2,1.95){$w_1 = 1$};
\node at (3.8,1.95){$w_2 = 0$};

%%%%%%%%%%%%%%%%%%%%%%%%%%%%%%%%%%%%
%% fill 2
%%%%%%%%%%%%%%%%%%%%%%%%%%%%%%%%%%%%
\node[anchor=west] at (0,0.3){fill = 2};
% suporte
\draw[line width=2pt] (1.5,0.3)--(4,0.3);
% curva
\draw[very thick,smooth,domain=1.5:4,samples=100]
plot (\x,{1/sqrt(2*3.1415*0.5^2)*exp(-(\x-2)^2/(2*0.5^2))+0.6});
% divisão
\draw[dashed] (3,-0.2)--(3,0.7);
% área esquerda
\fill[gray!20] (1.5,0.3) --  plot[smooth,domain=1.5:3,samples=100]
(\x,{1/sqrt(2*3.1415*0.5^2)*exp(-(\x-2)^2/(2*0.5^2)) + 0.6}) -- (3,0.3) -- cycle;
% área direita
\fill[gray!45] (3,0.3) -- plot[smooth,domain=3:4,samples=100]
 (\x,{1/sqrt(2*3.1415*0.5^2)*exp(-(\x-2)^2/(2*0.5^2)) + 0.6}) -- (4,0.3) -- cycle;
% desenha novamente a curva sobre o preenchimento
\draw[very thick,smooth,domain=1.5:4,samples=100]
 plot (\x,{1/sqrt(2*3.1415*0.5^2)*exp(-(\x-2)^2/(2*0.5^2))+0.6});
\draw[dashed, line width=2pt] (3,0.3)--(3,0.7);
\node at (2.35,-0.2){$w_1$};
\node at (3.45,-0.2){$w_2$};
\end{scope}
\end{tikzpicture}
\caption{Illustration of the three missing-data strategies implemented in \pkg{PRTree}. (a) Tree structure. (b) Corresponding partition of the predictor space. (c) Redistribution of the association probability for the three values of \code{fill\_type}.}
\label{fig:missing_tree}
\end{figure}

\subsection{Parameter estimation}\label{sec:parameter}
Given a training sample $\{(\bs X_i,Y_i)\}_{i=1}^{n}$, parameter estimation aims at minimizing the empirical quadratic loss
\begin{equation}\label{eq:tree}
\argmin_{\Theta \in \Xi}\Biggl\{\sum_{i=1}^{n} \biggl( Y_{i} - \sum_{m=1}^{M} \gamma_m P_{im} \biggr)^2\Biggr\},
\end{equation}
where $P_{im}=\Psi^\ast(\bs X_i;\Rg_m,\bs{\sigma})$ and the parameter space is
\[
\Xi=\Bigl\{(\{\Rg_m\}_{m=1}^{M},\bs\gamma,\bs{\sigma}):M\in\N\backslash\{0\},\ \Rg_m\subseteq\R^p,\ \bs\gamma\in\R^M,\ \bs{\sigma}\in\R_+^p\Bigr\}.
\]
The estimation of the different parameters in $\Theta$ alternates in between region and weight estimates, as in standard regression trees, until a stopping criterion is met. During this process, the number of regions is increased by one at each loop and the matrix $P$ and the weights $\bs \gamma$ are gradually updated. The vector $\bs{\sigma}$ can either fixed or be estimated through a grid search on a validation set.

\paragraph{Estimating $\bs{\gamma}$.} Given the selected smoothing vector $\bs{\sigma}$ and the terminal regions $\{\Rg_m\}_{m=1}^{M}$, the weight vector $\bs{\gamma}$ is estimated by solving the least-squares problem
\begin{equation}\label{eq:minimize_PRTree}
\hat{\bs{\gamma}}=\argmin_{\bs{\gamma}\in\R^M}\Biggl\{\sum_{i=1}^{n}\,\biggl(Y_i-\sum_{m=1}^{M}\gamma_mP_{im}\biggr)^2\Biggr\}=\argmin_{\bs{\gamma}\in\R^M}\bigl\{\bigl\|\bs Y-P\bs\gamma\bigr\|^2\bigr\},
\end{equation}
where $\bs Y=(Y_1,\cdots,Y_n)'$ and $P=(P_{im})$ is the matrix of association probabilities.

\paragraph{Estimating $\{\Rg_m\}_{m=1}^{M}$.} Assume that $M$ regions, referred to as current regions, have already been identified, meaning that the current tree has $M$ leaves. As in standard regression trees, each current region $\Rg_m$, $1 \leq m \leq M$, can be decomposed into two sub-regions $\Rf*{L}$ and $\Rf*{R}$ with respect to a coordinate $1 \leq j \leq p$ and a splitting point $t$ (threshold) that minimizes \eqref{eq:tree}. Each possible split leads to the update of $P$, that now belongs to $\R^{n\times (M+1)}$ (the space of $n$ by $M+1$ real matrices), and $\bs{\gamma}$, that now belongs to $\R^{M+1}$. The corresponding $\hat{\bs{\gamma}}$ is obtained through \eqref{eq:minimize_PRTree}. At each step, the split that leads to the smallest mean square error is selected.

\paragraph{Stopping.} The process terminates when any stopping criteria (e.g., depth, minimum node size) is met.

\section{The PRTree Package}\label{sec:package}
\subsection{Architecture}
The \pkg{PRTree} package implements the methodology described in Section~\ref{sec:prtree_method} within the \proglang{R} environment. It provides a unified interface for fitting Probabilistic Regression Trees with complete or incomplete predictor data, together with tools for prediction, visualization, model selection, and cross-validation. The computational core responsible for recursive tree construction and split optimization is implemented in FORTRAN. Linear algebra operations rely on the BLAS and LAPACK libraries distributed with \proglang{R}, while probability distribution, random number generators, and other low-level numerical routines are accessed through \proglang{R}'s native C API. Data preprocessing, model management, and graphical utilities are implemented in \proglang{R}.

Although the computational core is implemented in FORTRAN, the package exposes a compact and consistent \proglang{R} interface following the standard \proglang{R} workflow. Models are fitted through the function \code{pr_tree()}, which returns an object of class \code{"prtree"}. Standard S3 methods are provided for prediction, visualization, printing, and summarization, allowing the package to integrate naturally with the usual \proglang{R} workflow. Most aspects of the fitting algorithm are controlled through dedicated control objects, allowing users to modify both the tree-growing and cross-validation procedures without changing the main fitting interface. Consequently, routine analyses require only a small number of high-level functions for model fitting, data partitioning, and cross-validation, together with the corresponding methods for prediction, visualization, and model diagnostics.

Figure~\ref{fig:algo} summarizes the computational workflow implemented in \pkg{PRTree}. The estimation procedure begins in \proglang{R}, where the user inputs are validated before calling the FORTRAN subroutine \code{pr_tree_fort}. After initializing the required data structures, the subroutine iterates over the candidate smoothing parameter vectors $\bs{\sigma}$. For each candidate, the subroutine \code{build_tree} recursively constructs a probabilistic regression tree by repeatedly identifying splittable nodes, generating candidate splits, evaluating the selected candidates, and updating the tree whenever an admissible split is found. Once the tree has been grown, its predictive performance is evaluated, and the current best model is updated whenever an improvement is observed. After all candidate smoothing parameter vectors have been evaluated, the selected model is returned to \proglang{R}, where it becomes available for prediction, visualization, and diagnostic analyses. The following subsections describe the main user-level functions and control objects supporting this workflow.

\begin{figure}[!ht]
\centering
\begin{tikzpicture}
% Styles with align=center to allow \\ in nodes
\tikzstyle{startstop} = [rectangle, rounded corners, minimum width=2cm, minimum height=1cm, text centered, draw=black, fill=red!30, align=left]
\tikzstyle{process} = [rectangle, minimum width=2cm, minimum height=1cm, text centered, draw=black, fill=blue!20, align=left]
\tikzstyle{subrout} = [rectangle, minimum width=3cm, minimum height=1cm, text centered, draw=black,  align=left]
\tikzstyle{decision} = [rectangle, minimum width=3cm, minimum height=1cm, text centered, draw=black, fill=green!20, align=left]
\tikzstyle{arrow} = [thick,->,>=stealth]
% R -> Fortran
\node (start) [startstop, align = left] {
 In R: \code{pr_tree} validates the inputs and calls \code{pr_tree_fort}.\\[0.2cm]
 In FORTRAN: Initialize tree structures and workspace.};

 % Loop in sigma:
\node (selectsigma) [process, below of=start, anchor=north west] at (start.south west){
 \code{pr_tree_fort:}\\
 Loop over $\bs{\sigma}$ values.\\[0.2cm]
 Any candidates remaining?};
\draw [arrow] ([xshift=1cm]start.south west) -- ([xshift=1cm]selectsigma.north west);

% Subprocess for build_tree
\node (growtree) [process, right of = selectsigma, anchor=north west,
                  text width = 8cm, align=left, xshift = 1cm] at (selectsigma.north east) {
\code{build_tree}:\\
Repeat until a stopping criterion is met:
\begin{enumerate}[label=(\alph*)]
\item Identify splittable nodes.
\item Generate candidate splits (stage 1).
\item Evaluate candidate splits (stage 2).
\item Update the tree with the best split.
\end{enumerate}};
\draw[arrow] (selectsigma.north east) to[out=0,in=180] node[pos=0.6, above] {\textbf{YES}} (growtree.north west);

 % evaluate MSE
\node (evaluatetree) [decision, below of=growtree, anchor = north east] at (growtree.south east) {
Evaluate the current tree\\ and compute the test MSE\\[0.2cm]
Lower test MSE than current best?};
\draw[arrow] ([xshift=-2cm]growtree.south east) -- ([xshift=-2cm]evaluatetree.north east);

\node (updatetree) [process, left of=evaluatetree, anchor = south east, xshift = -1cm] at (evaluatetree.south west) {Update the best\\ tree status};
\draw[arrow] (evaluatetree.south west) to[out=180,in=0] node[pos=0.5, above] {\textbf{YES}} (updatetree.south east);

\coordinate (sigmaloop) at (selectsigma.east);
\draw[arrow] (updatetree.north) |- (sigmaloop);
\draw[arrow] (evaluatetree.north west) -- ++(-3.5,0) node[midway, above] {\textbf{NO}} |- (sigmaloop);

% End of loop
\node (end) [startstop, below of = selectsigma, anchor=north west] at (selectsigma.south west) {
    In FORTRAN: Finish the\\
    loop, process the output\\
    and return to R.\\[0.2cm]
    In R: Process the output\\
    and return the results};
\draw[arrow] (selectsigma) to[out=270,in=90] node[midway, left=3pt] {\textbf{NO}} ([xshift=3pt]end.north);
\end{tikzpicture}
\caption{Process flow for building a tree using the PRTree package.}
\label{fig:algo}
\end{figure}

\subsection{Main Interface}\label{sec:main}

Probabilistic Regression Trees are fitted through the function \code{pr_tree()}, which provides the main interface to the package. The function follows the standard modeling conventions of \proglang{R}: it receives the predictor data, the response variable, and an optional control object specifying the fitting procedure, returning an object of class \class{prtree}. This object can subsequently be analyzed through methods such as \code{predict()}, \code{plot()}, \code{print()}, and \code{summary()}.

The behavior of the fitting algorithm is controlled through the function \code{pr_tree_control()}, which encapsulates the algorithmic parameters governing tree construction, smoothing parameter selection, candidate split generation, missing-data handling, and computational options. Separating the model specification from the algorithmic configuration keeps the main interface compact while allowing detailed control of the estimation procedure whenever needed. Complete examples illustrating the use of \code{pr_tree()} and \code{pr_tree_control()} are provided in Section~\ref{sec:examples}.

The basic syntax of the fitting function is
\begin{Code}
pr_tree(X, y, control = pr_tree_control(),...)
\end{Code}
where \code{X} is the predictor matrix or data frame, \code{y} is the response vector, and \code{control} is a list of control parameters governing the fitting procedure. As a convenience, these parameters may also be supplied directly through \code{...}. Any remaining arguments are forwarded to the selected association density whenever required.

The control object is conveniently created through \code{pr_tree_control(...)}, which returns a list containing the algorithmic settings used during model fitting. The available control parameters can be grouped according to their role in the estimation procedure.
\begin{itemize}
\item \textbf{Smoothing parameters}, defining the candidate values of the smoothing vector $\bs{\sigma}$: \code{sigma_grid}, \code{grid_size}, \code{min_mult}, \code{max_mult}, and \code{tiny_sigma}.

\item \textbf{Tree complexity}, controlling tree growth through size, impurity reduction, and probability constraints: \code{max_terminal_nodes}, \code{max_depth}, \code{cp}, \code{n_min}, \code{prop_x}, and \code{p_min}.

\item \textbf{Validation settings}, defining the hold-out sample used for smoothing-parameter selection: \code{prop_hold} and \code{idx_train}.

\item \textbf{Missing-data handling}, specifying the strategy for evaluating association probabilities and the proxy criterion adopted during candidate generation: \code{fill_type} and \code{proxy_crit}.

\item \textbf{Candidate split search}, controlling the first stage of the split-selection procedure: \code{n_candidates} and \code{by_node}.

\item \textbf{Association density}, specifying the probability density used to compute association probabilities through \code{dist} together with any distribution-specific parameters supplied via \code{...}.

\item \textbf{Output and execution}, controlling printed output and execution messages through \code{iprint} and \code{verbose}.
\end{itemize}
When called without arguments, the function returns a control object initialized with the default values:

\begin{CodeChunk}
\begin{CodeInput}
R> pr_tree_control()
\end{CodeInput}
\begin{CodeOutput}
========================================================
PRTree Control Parameters
========================================================
Sigma grid: not provided (will be generated with size 8)
Multiplier range: [0, 2]
Validation set proportion: 0.20
Max terminal nodes: 15
Max depth: 14
Complexity parameter (cp): 0.010
Minimum observations per node: 5
Proportion threshold (prop_x): 0.10
Probability threshold (p_min): 0.050
Fill type: 2
Proxy criterion: both
Number of candidates: 3
By node: FALSE
Distribution: norm
========================================================
\end{CodeOutput}
\end{CodeChunk}

\subsection{Association Densities}

The implementation of \pkg{PRTree} assumes that the multivariate association density factorizes as the product of its univariate marginal densities. More precisely,
\[
\phi(\bs{z})=\prod_{j=1}^{p}\phi(z_j).
\]
The same symbol $\phi$ is used to denote both the multivariate association density and its univariate marginal density. This slight abuse of notation is unambiguous because the joint density is completely determined by its marginals, all of which belong to the same parametric family. Consequently, restricting the association density to any subset of coordinates preserves the same family of distributions. Consequently, the association probability in Equation~\eqref{eq:Psi} is evaluated as
\[
\Psi(\bs{X};\Rg,\bs{\sigma})=\int_{\Rg}\prod_{j=1}^{p}\frac{1}{\sigma_j}\phi\biggl(\frac{v_j-X_j}{\sigma_j}\biggr)d\bs{v}.
\]

The marginal density is specified through the argument \code{dist}, while any distribution-specific parameters are supplied through \code{dist_pars}. The choice of the association density and its distribution-specific parameters is part of the model specification. In practice, these quantities are typically selected jointly with the smoothing parameter $\bs{\sigma}$ through cross-validation using a predictive performance criterion. The package currently supports the following parameterizations.

\paragraph{Gaussian.} The Gaussian association density requires no additional parameters:
\[
\phi(z)=\frac{1}{\sqrt{2\pi}}\exp\biggl\{-\frac{z^2}{2}\biggr\}.
\]

\paragraph{Student's \texorpdfstring{$t$}{t}.} The Student's $t$ association density is parameterized by its degrees of freedom $\nu>0$, specified through \code{dist\_pars = list(df = ...)}:
\[
\phi(z)=\frac{\Gamma\bigl((\nu+1)/2\bigr)}{\Gamma(\nu/2)\sqrt{\pi\nu}}\biggl(1+\frac{z^2}{\nu}\biggr)^{-(\nu+1)/2}.
\]

\paragraph{Log-normal.} The Log-normal association density is parameterized by its logarithmic standard deviation $s>0$, specified through \code{dist\_pars = list(sdlog = ...)}:
\[
\phi(z)=\frac{1}{zs\sqrt{2\pi}}\exp\biggl\{-\frac{[\log(z)]^2}{2s^2}\biggr\}I(z>0).
\]

\paragraph{Gamma.} The Gamma association density is parameterized by its shape parameter $\alpha>0$, specified through \code{dist\_pars = list(shape = ...)}:
\[
\phi(z)=\frac{z^{\alpha-1}e^{-z}}{\Gamma(\alpha)}I(z>0).
\]

In all cases, the predictor value $X_j$ acts as the location (or lower bound for asymmetric distributions), while the smoothing parameter $\sigma_j$ controls the scale of the association density through the transformation $z=(v_j-X_j)/\sigma_j$. Consequently, the user only specifies the remaining shape parameters, when applicable.

\subsection{Auxiliary Functions}

Besides the main fitting interface, \pkg{PRTree} provides a collection of auxiliary functions supporting model construction, resampling, prediction, visualization, and result inspection. These functions can be used independently or combined into customized workflows, while remaining fully compatible with the main fitting functions.

\subsubsection*{Sigma Grid Generation}

The function \code{expand_sigma_grid()} generates the grid of candidate smoothing vectors used by \code{pr_tree()} whenever the argument \code{sigma_grid} is not supplied. The construction of the grid proceeds as follows:

\begin{enumerate}
\item The sample standard deviation $\hat s_j$ is computed for each predictor using the training sample.

\item The candidate vectors
\[
\bs{\sigma}^{(k)}=c_k(\hat s_1,\cdots,\hat s_p),\quad k=1,\cdots,K,
\]
are generated, where $K=\text{\code{grid\_size}}$ and $c_1,\cdots,c_K$ are equally spaced multipliers determined by the arguments \code{min_mult} and \code{max_mult}. By default, \code{grid_size = 8}, \code{min_mult = 0}, and \code{max_mult = 2}.
\end{enumerate}

Optionally, the argument \code{tiny_sigma} may be used to prepend an additional candidate to the generated grid. Setting \code{tiny_sigma = 0} includes the deterministic case $\bs{\sigma}=\bs{0}$, for which the probabilistic assignment coincides exactly with the hard splitting rule of classical regression trees. Positive values of \code{tiny_sigma} provide smooth approximations of this limiting case.

Although this procedure is performed automatically whenever \code{sigma_grid} is not supplied, \code{expand_sigma_grid()} is also available as a user-level function, allowing the generated grid to be inspected or customized before fitting the model. When the predictor matrix is not available, supplying only \code{n_feat} generates the corresponding default grid based on unit feature scales, which can subsequently be modified or passed directly to \code{pr_tree()}.

\begin{CodeChunk}
\begin{CodeInput}
R> set.seed(1234)
R> expand_sigma_grid(n_feat = 3)
R> expand_sigma_grid(X = matrix(runif(100 * 3, 0, 10), ncol = 3))
\end{CodeInput}
\begin{CodeOutput}
========================================================
Sigma Grid
========================================================
Grid size: 8 candidates
Features: 3

First 5 rows:
       V1   V2   V3
[1,] 0.25 0.25 0.25
[2,] 0.50 0.50 0.50
[3,] 0.75 0.75 0.75
[4,] 1.00 1.00 1.00
[5,] 1.25 1.25 1.25
... (3 more rows)
========================================================

========================================================
Sigma Grid
========================================================
Grid size: 8 candidates
Features: 3

First 5 rows:
         V1     V2    V3
[1,] 0.6968 0.7321 0.663
[2,] 1.3935 1.4641 1.326
[3,] 2.0903 2.1962 1.989
[4,] 2.7870 2.9282 2.652
[5,] 3.4838 3.6603 3.315
... (3 more rows)
========================================================
\end{CodeOutput}
\end{CodeChunk}

\subsubsection*{Data Splitting}

The function \code{train_test_split()} generates data partitions for training, validation, and cross-validation. When \code{n_rep = NULL} (default), it creates a single training/test split. Alternatively, setting \code{n_rep} generates a collection of folds suitable for $K$-fold cross-validation. In both cases, the function supports either simple random sampling or stratified sampling based on the presence of missing values, ensuring that the proportion of incomplete observations is approximately preserved across the generated subsets. The returned object stores the corresponding indices together with summary information describing the resulting partition. An example illustrating the generation of cross-validation folds is provided in Section~\ref{sec:examples}.

The following example compares simple random sampling with stratified sampling. In this particular realization, simple random sampling allocates all observations with missing values to the training set, whereas stratification preserves the overall proportion of incomplete observations in both subsets.

\begin{CodeChunk}
\begin{CodeInput}
R> set.seed(1234)
R> n_obs <- 100
R> is_NA <- c(rep(TRUE, 10), rep(FALSE, 90))
R> summary(train_test_split(n_obs = n_obs, is_NA = is_NA))
R> summary(train_test_split(n_obs = n_obs, is_NA = is_NA, stratify = TRUE))
\end{CodeInput}
\begin{CodeOutput}
========================================================
Train/Test Split Summary
========================================================
Sampling method: Simple random

Note: Missing values statistics are shown below but were NOT
      used for stratification (stratify = FALSE).

Total observations: 100
Training set: 80 (80.0%)
Testing set: 20 (20.0%)

--- Missing Values ---
Training set: 10 (12.5%)
Testing set:  0 (0.0%)
Overall:      10 (10.0%)
========================================================

========================================================
Train/Test Split Summary
========================================================
Sampling method: Stratified (by missing values)
Total observations: 100
Training set: 80 (80.0%)
Testing set: 20 (20.0%)

--- Missing Values ---
Training set: 8 (10.0%)
Testing set:  2 (10.0%)
Overall:      10 (10.0%)
========================================================
\end{CodeOutput}
\end{CodeChunk}

\subsubsection*{Cross-Validation}

Cross-validation is implemented through \code{pr_tree_cv()}, while resampling-specific settings are specified using \code{pr_tree_control_cv()}, which extends \code{pr_tree_control()} with cross-validation parameters. The resampling scheme is selected through the argument \code{method}, with available options \code{"montecarlo"} (Monte Carlo) and \code{"kfold"} ($K$-fold). For $K$-fold cross-validation, users may optionally provide the argument \code{fold_idx}, specifying the fold assignment of each observation. This allows identical data partitions to be reused across different models or experimental settings, ensuring fair performance comparisons.

The framework supports three common workflows.

\begin{enumerate}[leftmargin=*]
\item \textbf{Joint smoothing-parameter selection and prediction-error estimation (default).} Candidate smoothing vectors are evaluated within each resampling iteration, the best-performing vector is selected, and its predictive performance is subsequently estimated on an independent test set.

\item \textbf{Smoothing-parameter selection only.} Setting \code{only_sigma = TRUE} omits the test stage, allowing the resampling procedure to focus exclusively on selecting the smoothing parameter. Repeating the procedure across multiple resampling iterations also provides information about the stability of the selected smoothing vectors.

\item \textbf{Prediction-error estimation for a fixed smoothing parameter.} When the supplied \code{sigma_grid} contains a single candidate, smoothing-parameter selection is skipped and cross-validation is used only to estimate the predictive performance of that fixed configuration.
\end{enumerate}

Under the default workflow, each resampling iteration first partitions the data into training and test sets. For \code{method = "kfold"}, the test set consists of one fold and the remaining folds form the training set. For \code{method = "montecarlo"}, the training and test sets are generated by random sampling according to the proportion specified by \code{prop_test}.

The training set may then be further divided into training and validation subsets according to \code{prop_valid}. When \code{prop_valid > 0}, each candidate smoothing vector is fitted on the training subset and evaluated on the validation subset. The smoothing vector minimizing the validation MSE is selected, after which the model is optionally refitted on the complete training set before evaluating its predictive performance on the corresponding test set. This final refitting step is controlled by the argument \code{update_final}. When \code{update_final = FALSE}, the model fitted on the reduced training subset is evaluated directly on the test set. When \code{prop_valid = 0}, no validation subset is created and candidate smoothing vectors are instead compared using their training MSE computed on the complete training set.

The returned object stores the selected smoothing vectors together with the corresponding validation and test prediction errors whenever these quantities are computed. These results can be used to compare candidate smoothing parameters, estimate predictive performance, assess the stability of the selected smoothing vectors across different resampling iterations, and identify an appropriate smoothing configuration for fitting the final model.

\subsubsection*{Prediction}

Predictions for new observations are obtained through the standard S3 generic \code{predict()}. The method accepts matrices or data frames containing new predictor values, including observations with missing covariates. By default, the function returns the predicted responses. Setting \code{complete = TRUE} additionally returns the corresponding association probability matrix, allowing users to inspect how each observation is probabilistically associated with the terminal regions.

\subsubsection*{Visualization and Summaries}

The package implements standard S3 methods for model inspection. The generic functions \code{print()} and \code{summary()} provide concise and detailed summaries for fitted trees, control objects, cross-validation results, and training/test partitions.

Visualization is provided through two complementary interfaces. For fitted trees, the S3 generic \code{plot()} produces diagnostic graphics, including fitted-versus-observed plots, residual diagnostics, and summaries of the association probabilities in the terminal regions. The dedicated function \code{plot_tree()} provides a graphical representation of the fitted tree itself, displaying the splitting rules, the estimated terminal-node weights, and compact heatmaps summarizing the association probabilities within each terminal region. For cross-validation objects, \code{plot()} summarizes the resampling results through graphics of the validation and test prediction errors, the selected smoothing vectors across iterations or folds, their empirical distributions, and their relationship with the validation error.

\subsection{Tree Construction in FORTRAN}

The subroutine \code{pr_tree_fort} serves as the interface between R and the FORTRAN implementation and coordinates the entire tree-construction procedure, whose main workflow is summarized in Figure~\ref{fig:algo}. First, it initializes the tree structure and allocates the output objects. It then verifies whether the root node admits at least one candidate threshold; otherwise, the algorithm immediately returns the root-only tree. Next, the routine iterates over the candidate values of the smoothing parameter $\bs{\sigma}$. For each value, it rebuilds the root state, constructs the corresponding tree by calling the recursive subroutine \code{build_tree}, computes the associated predictions, and compares the resulting prediction error with the best solution obtained so far. Whenever a better tree is found, the corresponding model is retained and, in full-output mode, serialized to the output objects. After all candidate values of $\bs{\sigma}$ have been evaluated, the routine returns the selected tree together with the requested predictions and summary statistics.
% What \code{pr_tree_fort} does:
% \begin{enumerate}[label=\arabic*.]
% \item Initialize the tree model (\code{set_tree_model}, \code{assign_pointers}, \code{start_root}).
% \item Check whether the root node contains admissible thresholds. If not, return the root-only tree (\code{return_root_only_tree}).
% \item For each candidate value of $\bs{\sigma}$: rebuild the root state (\code{start_root}); build the tree (\code{build_tree}); compute the predictions (\code{predict_tree}); compare the prediction error with the best solution obtained so far; whenever the current tree is preferred, update the stored best model and, in full-output mode, serialize it through \code{return_tree}.
% \item Return the selected tree together with the requested predictions and summary statistics.
% \end{enumerate}

The recursive subroutine \code{build_tree} is responsible for constructing the tree associated with a fixed value of the smoothing parameter $\bs{\sigma}$. Starting from the root node, the algorithm repeatedly identifies the terminal nodes eligible for further partitioning, updates the information required for split evaluation, searches for promising candidate splits through a two-stage procedure, and incorporates the best candidate into the tree whenever it satisfies the complexity criterion. This process continues until no further admissible split can be found, the limit imposed by \code{max_terminal_nodes} is reached, or another stopping criterion is satisfied. The following subsections describe the key components of this procedure in detail.

\subsubsection{Node Eligibility Criteria}

Before each iteration of the tree-growing procedure, the algorithm updates the list of terminal nodes eligible for further partitioning. Only newly created child nodes are evaluated against the eligibility criteria described in this subsection. Existing splittable nodes are retained unless they have been declared terminal during the tree-growing procedure.

The argument \code{max_depth} specifies the maximum depth allowed for any terminal node, where the depth is defined as the number of edges on the path from the root node. Child nodes reaching this limit are immediately declared terminal. The remaining newly created child nodes are then screened according to the arguments \code{prop_x} and \code{p_min}.

Since the posterior probability of a parent region equals the sum of the posterior probabilities of its two child regions, a node with insufficient posterior probability mass cannot produce two child regions simultaneously satisfying the criterion defined by \code{prop_x} and \code{p_min}. Moreover, attempting to split such nodes would unnecessarily increase the size of the association probability matrix by introducing nearly zero columns, thereby increasing the computational cost of subsequent iterations and potentially deteriorating the numerical conditioning of the least-squares problem.

The implementation therefore requires that at least a proportion \code{prop_x} of the training observations satisfy $P^\ast(\Rg|\bs X_i)\ge2\times\text{\code{p\_min}}$, where $\Rg$ denotes the current terminal region. This provides a sufficient screening criterion, since any node failing this condition cannot produce two child regions simultaneously satisfying the criterion defined by \code{prop_x} and \code{p_min}. Satisfying this preliminary criterion does not, however, guarantee that the resulting child nodes will also satisfy the probability constraint, as this ultimately depends on the candidate threshold.

\subsubsection{Threshold Computation}

For each node remaining after the eligibility screening, the algorithm updates the information required for split evaluation. This includes computing the admissible candidate thresholds for every predictor and updating the set of important features inherited from the parent node.

The argument \code{n_min} specifies the minimum number of observations required in each child node produced by a candidate split. Thresholds are computed separately for each predictor using only the non-missing observations for that predictor. Consequently, observations routed through the tree with a missing value for the current predictor are ignored when generating thresholds, although they remain associated with the node and may contribute to threshold computation for other predictors.

Unlike classical regression trees, probabilistic terminal regions need not contain any observations, since observations are associated with regions through posterior probabilities rather than hard assignments. Nevertheless, the implementation follows the standard recursive-partitioning literature and generates candidate thresholds as the midpoints between consecutive distinct observed values of the candidate predictor. Restricting the search to these midpoints preserves every threshold at which the induced partition changes while substantially reducing the computational cost of the search.

To satisfy the minimum-node-size constraint, threshold computation is attempted only when at least $2\times\text{\code{n\_min}}$ non-missing observations are available for the current predictor. After sorting the observed values, duplicated values are removed and candidate thresholds are defined as the midpoints between consecutive distinct values. If no admissible threshold is found for any predictor, the node is immediately declared terminal.

Threshold generation only enforces the minimum-node-size constraint through the argument \code{n_min}. The probability criterion defined by \code{prop_x} and \code{p_min} is evaluated later during tree construction, since verifying compliance of the resulting child nodes requires computing their posterior probabilities. Candidate thresholds failing this criterion are discarded before entering the proxy evaluation stage. The remaining thresholds are passed to Stage~1.

\subsubsection*{Candidate Search -- Stage 1}

For each eligible terminal node, Stage~1 searches for the most promising candidate splits. The search is performed independently for every predictor with at least one admissible threshold. Rather than scanning the candidate thresholds sequentially, the implementation starts from the center of the threshold list and progressively explores both directions. This strategy allows the search to terminate early whenever the probability constraints can no longer be satisfied. Since the candidate thresholds and their proxy scores depend only on the observations associated with the current terminal node, this search is performed only once for each node. The resulting candidate list is reused in subsequent iterations until the node is eventually split or declared terminal.

Although the terminal-node weights ($\gamma_m$'s) and the corresponding empirical loss are not recomputed during Stage~1, the posterior probabilities of the child regions must still be evaluated to verify whether a candidate threshold satisfies the admissibility criterion defined by \code{prop_x} and \code{p_min}. While this introduces additional computations compared with classical regression trees, it remains substantially less expensive than the complete probabilistic evaluation performed during Stage~2 because it does not require re-estimating the terminal-node weights. Moreover, postponing this verification until Stage~2 could allow inadmissible candidate thresholds to be selected by the proxy score, potentially excluding admissible candidates from the fixed-size candidate list passed to the second stage. Candidate thresholds failing the probability criterion are therefore discarded before proxy ranking.

The remaining candidates are then ranked using the proxy score. For node $m$, predictor $j$, and candidate threshold $t$, the score is defined as
\[
S^{(m)}_{j,t} = \frac{\bigl(\sum_{i\in\mathcal{I}_L(j,t)}Y_i\bigr)^2}{|\mathcal{I}_L(j,t)|} + \frac{\bigl(\sum_{i\in\mathcal{I}_R(j,t)}Y_i\bigr)^2}{|\mathcal{I}_R(j,t)|},
\]
where $\mathcal{I}_L(j,t)$ and $\mathcal{I}_R(j,t)$ denote the sets of indices corresponding to the observations assigned to the left and right child nodes, respectively. For candidate splitting variables containing missing values, these assignments are completed using the proxy routing procedure described in the sequel. In the absence of missing values, these sets reduce to $\mathcal{I}_L(j,t)=\{i:X_{ij}\le t\}$ and $\mathcal{I}_R(j,t)=\{i:X_{ij}>t\}$.

This score coincides with the splitting criterion used in classical regression trees and provides a computationally inexpensive approximation to the reduction in empirical loss. Although it ignores the probabilistic association matrix and the re-estimation of the terminal-node weights, it provides an effective heuristic for identifying the most promising candidate splits while avoiding the cost of repeatedly fitting the complete probabilistic model.

The proxy score is used exclusively as a screening criterion during Stage~1. Only the highest-ranked candidates proceed to Stage~2, where they are evaluated using the complete probabilistic model. The number of retained candidates is controlled by the argument \code{n_candidates}. When \code{by_node = TRUE}, the best candidates are selected independently within each eligible node; otherwise, a single global ranking is constructed across all eligible nodes.

\subsubsection*{Handling Missing Values}\label{sec:missing_stage1}

When the candidate splitting variable contains missing values, the assignment of observations to the left and right child nodes is no longer uniquely determined by the candidate threshold. Stage~1 therefore employs a proxy routing procedure, similar in spirit to surrogate splits, to evaluate each candidate split. First, all observations with non-missing values for the candidate predictor are assigned to the left or right child node according to the candidate threshold. The remaining observations are then processed sequentially according to the criterion specified by \code{proxy_crit}. The resulting assignments are retained throughout the remaining tree-growing process whenever the corresponding split is ultimately selected.

Let $\mathcal{I}_L$ and $\mathcal{I}_R$ denote the current sets of observations assigned to the left and right child nodes, respectively, with cardinalities $n_L=|\mathcal{I}_L|$ and $n_R=|\mathcal{I}_R|$, and let $S_L=\sum_{i\in\mathcal{I}_L}Y_i$ and $S_R=\sum_{i\in\mathcal{I}_R}Y_i$ denote the corresponding response sums. Finally, let $\bs X_{\mathrm{miss}}$ denote an observation with a missing value for the candidate predictor and response $Y_{\mathrm{miss}}$. For each incomplete observation, scores $C_L$ and $C_R$ are computed for assigning the observation to the left or right child node. The observation is assigned to the node with the larger score, after which the corresponding node counts and response sums are immediately updated before the next incomplete observation is processed. This procedure is repeated until all incomplete observations associated with the candidate split have been assigned. Three proxy criteria are available.

\paragraph{\code{proxy\_crit = "mean"}.} Maximizes the separation between the child-node means,
\[
C_L=\biggl|\frac{S_L+Y_{\mathrm{miss}}}{n_L+1}-\frac{S_R}{n_R}\biggr|,\qquad
C_R=\biggl|\frac{S_R+Y_{\mathrm{miss}}}{n_R+1}-\frac{S_L}{n_L}\biggr|.
\]

\paragraph{\code{proxy\_crit = "var"}.} Maximizes between-node variance (proxy for the reduction in the residual sum of squares),
\[
C_L=\frac{(S_L+Y_{\mathrm{miss}})^2}{n_L+1}+\frac{S_R^2}{n_R},\qquad
C_R=\frac{(S_R+Y_{\mathrm{miss}})^2}{n_R+1}+\frac{S_L^2}{n_L}.
\]

\paragraph{\code{proxy\_crit = "both"}.} Uses the sum of the previous two scores.

The resulting assignments are not used solely to evaluate the current candidate split. Once a split is accepted, routed observations become permanently associated with their assigned child node and participate in all subsequent recursive splitting steps. Candidate thresholds are always generated using the observations with non-missing values for the current splitting variable. Therefore, a routed observation contributes to threshold computation whenever its value for the candidate predictor is observed and is ignored otherwise. After model fitting, the terminal node associated with each training observation is returned as part of the fitted model object. Together with the fitted tree structure, this information uniquely determines the complete path followed by every observation through the tree.

\subsubsection*{Complete Candidate Evaluation -- Stage~2}

The candidates retained from Stage~1 undergo a complete probabilistic evaluation. For each candidate split, the association probability matrix is recomputed, the terminal-node weights are re-estimated by solving the least-squares problem described in Section~\ref{sec:parameter}, and the corresponding empirical loss is evaluated. The least-squares problem is solved using the LAPACK routine \code{DGELSD}, which computes the minimum-norm solution through a divide-and-conquer singular value decomposition. This approach provides numerical stability and naturally accommodates rank-deficient association probability matrices.

After all retained candidates have been evaluated, the candidate producing the largest reduction in the empirical loss is identified. As in classical CART, the argument \code{cp} controls the trade-off between model fit and tree complexity by preventing splits that do not produce a sufficiently large improvement. Unlike the post-pruning strategy traditionally associated with CART, however, this criterion is enforced during tree construction. Let $\mathrm{MSE}_{\mathrm{curr}}$ denote the empirical loss of the current tree and $\mathrm{MSE}_{\mathrm{new}}$ the loss obtained after incorporating the candidate split. The selected split is accepted only if
\[
\mathrm{MSE}_{\mathrm{new}}\le(1-\text{\code{cp}})\,\mathrm{MSE}_{\mathrm{curr}}.
\]
Equivalently, the relative reduction in the empirical loss must exceed the threshold specified by \code{cp}. Otherwise, no further split is accepted and all remaining splittable nodes are declared terminal, terminating the tree-growing procedure. When a split is accepted, the corresponding child nodes are added to the tree, and the algorithm proceeds with the next iteration.

\section{Illustrative Examples}\label{sec:examples}

\subsection{Data description}

We illustrate the main functionalities of \pkg{PRTree} using the \code{economics} data set from the \pkg{ggplot2} package. The response variable is the median duration of unemployment (\code{uempmed}), while the remaining numeric variables are used as predictors. The predictors are extracted as a standard \code{data.frame}, which is the expected input format for \pkg{PRTree}.
\begin{CodeChunk}
\begin{CodeInput}
R> library("PRTree")
R> library("ggplot2")
R> data("economics")
R> X <- as.data.frame(economics[, sapply(economics, is.numeric)])
R> y <- X$uempmed
R> X$uempmed <- NULL
\end{CodeInput}
\end{CodeChunk}

Since this data set does not contain missing values, we additionally construct a modified version by artificially introducing missing values into the predictor variables. Specifically, predictor values exceeding the 90th percentile of each variable are set to missing, yielding a simple Missing Not At Random (MNAR) mechanism in which the probability of missingness depends on the value of the covariate itself. The resulting data set, denoted by \code{X\_miss}, is used throughout the following examples to illustrate the missing-data capabilities of \pkg{PRTree}.
\begin{CodeChunk}
\begin{CodeInput}
R> set.seed(1234)
R> X_miss <- X
R> for (i in 1:ncol(X)) {
+   q90 <- quantile(X_miss[, i], 0.9)
+   X_miss[X_miss[, i] > q90, i] <- NA
+ }
\end{CodeInput}
\end{CodeChunk}

\subsection{Basic usage}
Model fitting is controlled through a \code{pr_tree_control()} object. In this example, we use a Gamma association function with shape parameter 1.5, enable nodewise candidate selection, and limit the tree to at most 30 terminal nodes, while all remaining arguments are left at their default values.
\begin{CodeChunk}
\begin{CodeInput}
R> ctrl <- pr_tree_control(
+   max_terminal_nodes = 30,
+   dist = "gamma",
+   shape = 1.5,
+   by_node = TRUE)
\end{CodeInput}
\end{CodeChunk}

The model is fitted using \code{pr_tree()}. The resulting object is of class \code{"prtree"} and implements the standard S3 methods \code{print()}, \code{summary()}, \code{plot()}, and \code{predict()}. The \code{print()} method provides a concise overview of the fitted tree, reporting the model dimensions, the selected association function, the training, validation, and global mean squared errors, and the selected smoothing parameters.
\begin{CodeChunk}
\begin{CodeInput}
R> set.seed(1234)
R> fit <- pr_tree(y = y, X = X, control = ctrl)
R> print(fit)
\end{CodeInput}
\begin{CodeOutput}
========================================================
Probabilistic Regression Tree (PRTree)
========================================================
Number of observations: 574
Number of features: 4
Number of terminal nodes: 13
Distribution: gamma
  with shape = 1.5000

Mean Squared Error:
  Train: 0.650891
  Validation: 1.279054
  Global: 0.776743

Selected sigma:
   1787.8132, 18415.0312, 1.4961, 1357.1985
========================================================
\end{CodeOutput}
\end{CodeChunk}

The \code{summary()} method extends the information displayed by \code{print()} with a detailed summary of the terminal regions. For each terminal node, it reports the estimated terminal value (\code{gamma}) together with descriptive statistics of the corresponding association probabilities, including the mean, standard deviation, minimum, quartiles, median, and maximum. The returned object can also be used programmatically for further analyses.
\begin{CodeChunk}
\begin{CodeInput}
R> s <- summary(fit)
R> print(head(s$terminal_nodes), digits = 4, row.names = FALSE)
\end{CodeInput}
\begin{CodeOutput}
 node   gamma  P_mean    P_sd P_min    P_q25 P_median    P_q75  P_max
    1 -73.118 0.06023 0.06763     0 0.007335  0.02518 0.102671 0.2102
    2  49.655 0.08655 0.10650     0 0.008863  0.02945 0.155229 0.3239
    3  57.708 0.03977 0.07881     0 0.001531  0.01010 0.032083 0.3853
    4   5.262 0.09224 0.20511     0 0.000000  0.00000 0.002516 0.7082
    5  -1.451 0.04008 0.06591     0 0.000000  0.00000 0.065545 0.2451
    6  14.025 0.03979 0.06923     0 0.000000  0.00000 0.059235 0.2854
\end{CodeOutput}
\end{CodeChunk}

\begin{figure}[!ht]
\centering
\includegraphics[width=\textwidth]{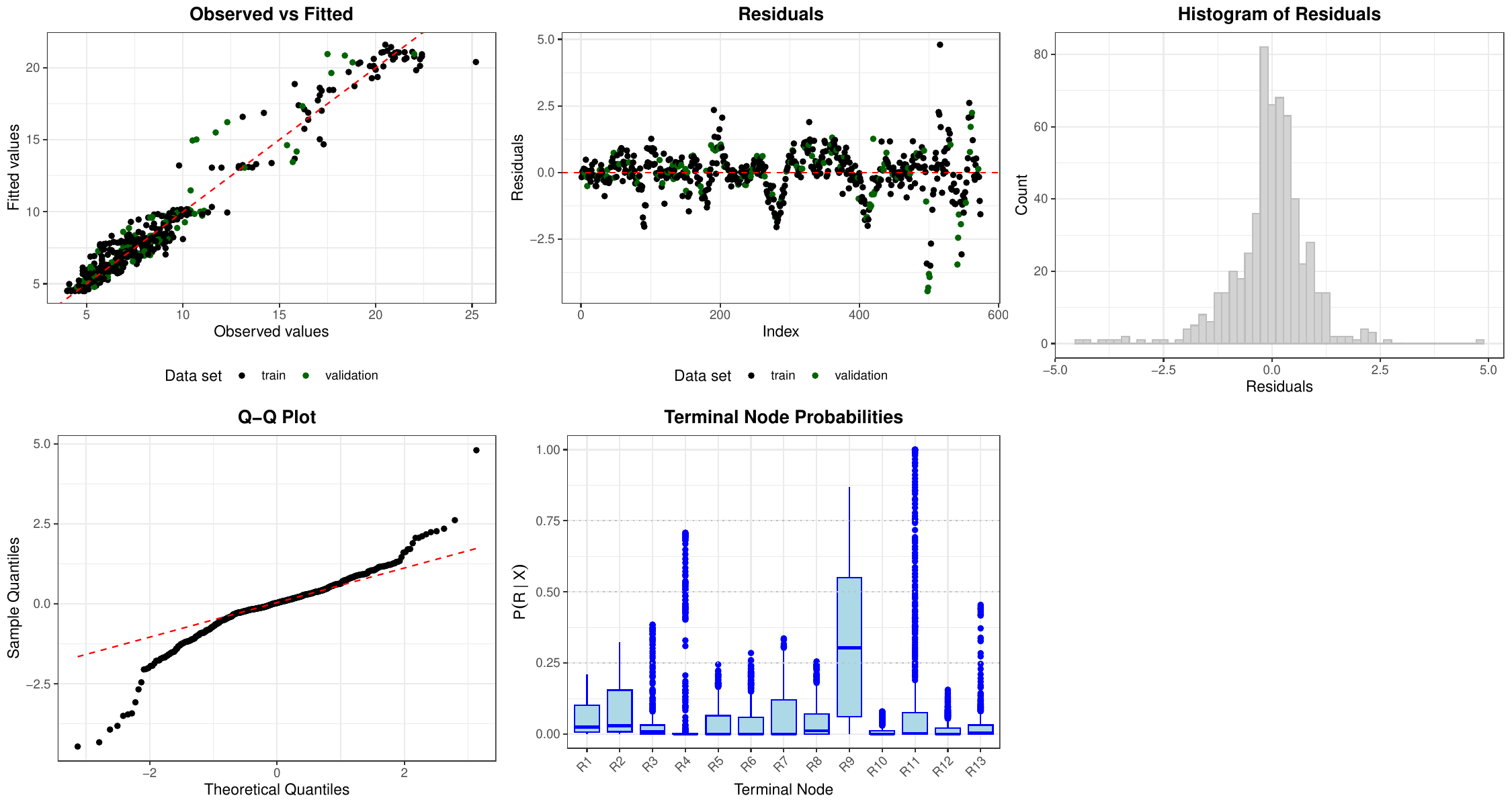}
\caption{\label{fig:diagnostics} Diagnostic plots produced by \code{plot()}. The first four panels evaluate the model fit through the observed versus fitted values, residuals by observation index, histogram of residuals, and normal Q--Q plot. The fifth panel summarizes the empirical distribution of the association probabilities for each terminal node, providing a graphical description of the probabilistic partition induced by the fitted tree.}
\end{figure}

The generic \code{plot()} method provides a collection of diagnostic plots for fitted probabilistic regression trees. By default, it produces five complementary graphics: observed versus fitted values, residuals by observation index, a histogram of the residuals, a normal Q--Q plot, and boxplots of the terminal-node association probabilities. Alternatively, individual plots can be displayed by specifying the corresponding indices through the \code{which} argument. The resulting diagnostic plots are displayed in Figure~\ref{fig:diagnostics}.
\begin{CodeChunk}
\begin{CodeInput}
R> plot(fit, y = y)
\end{CodeInput}
\end{CodeChunk}

\begin{figure}[!ht]
\centering
\includegraphics[width=\textwidth]{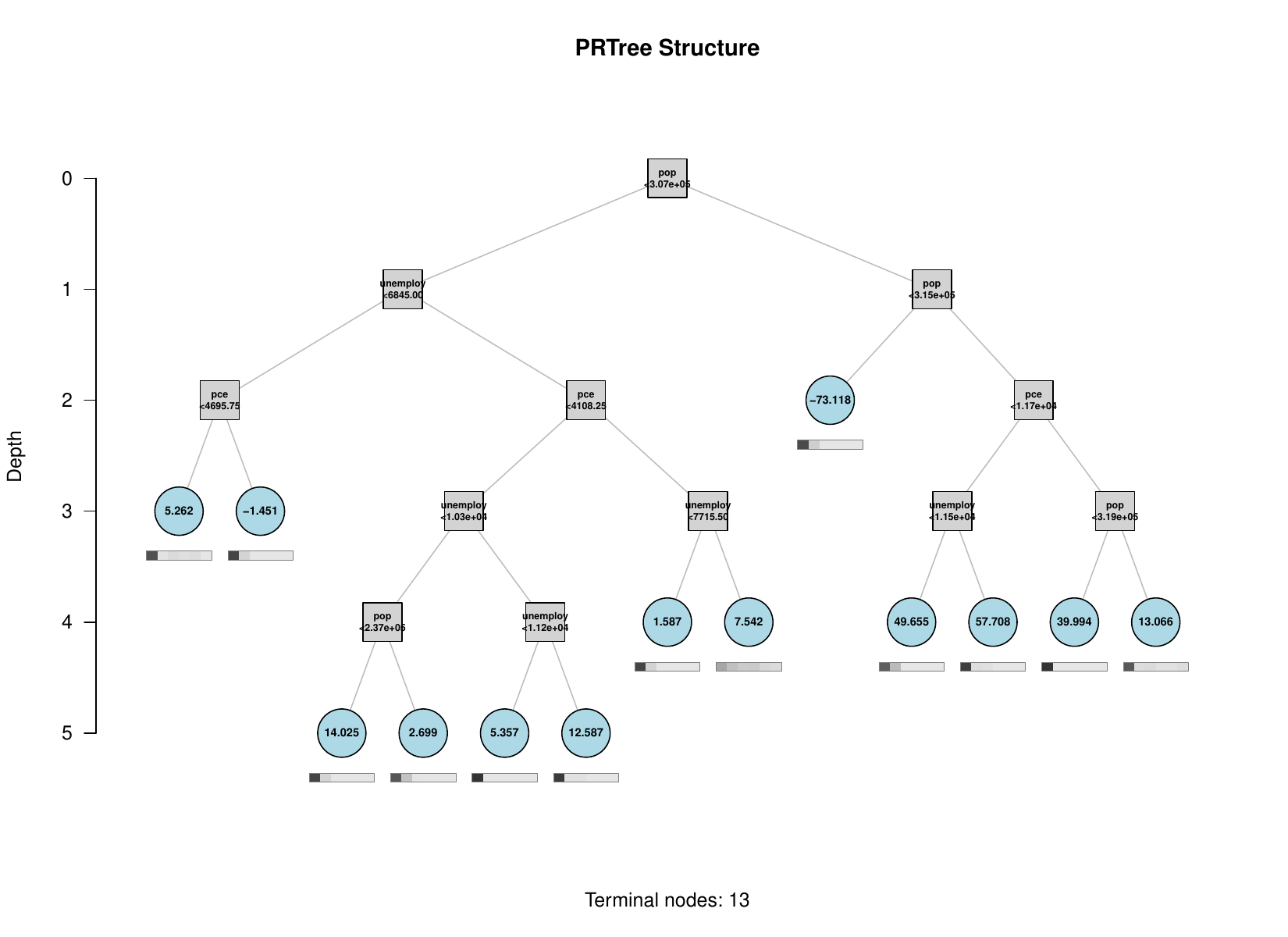}
\caption{\label{fig:tree} Tree representation produced by \code{plot\_tree()}. Internal nodes are labeled with the splitting variable and threshold, terminal nodes display the estimated terminal values (\code{gamma}), and the heatmaps summarize the empirical distribution of the association probabilities for each terminal region.}
\end{figure}

The \code{plot_tree()} function displays the fitted probabilistic regression tree as a dendrogram. Internal nodes are labeled with the selected splitting variable and threshold, while terminal nodes display the corresponding estimated terminal values (\code{gamma}). Additionally, a mini heatmap is shown below each terminal node, summarizing the empirical distribution of the corresponding association probabilities. This visualization provides an intuitive representation of both the tree structure and the probabilistic partition induced by the model. The resulting tree representation is shown in Figure~\ref{fig:tree}.
\begin{CodeChunk}
\begin{CodeInput}
R> plot_tree(fit)
\end{CodeInput}
\end{CodeChunk}

Predictions for new observations are obtained through the generic
\code{predict()} method.

\begin{CodeChunk}
\begin{CodeInput}
R> head(predict(fit, X))
\end{CodeInput}
\begin{CodeOutput}
[1] 4.671657 4.668496 4.668894 4.756089 4.713778 4.685151
\end{CodeOutput}
\end{CodeChunk}

\subsection{Handling Missing Predictors}

A distinctive feature of \pkg{PRTree} is its ability to fit probabilistic regression trees directly from data sets containing missing predictor values, without requiring a separate imputation step. Since the smoothing parameters are selected using an internal validation sample, we first generate a training/validation partition using \code{train_test_split()}. In this example, the partition is stratified according to the presence of missing values, ensuring that both subsets contain similar proportions of incomplete observations. The resulting partition is used only by \pkg{PRTree} to select the smoothing parameters.

\begin{CodeChunk}
\begin{CodeInput}
R> set.seed(1234)
R> idx <- train_test_split(
+   n_obs = nrow(X_miss),
+   is_NA = apply(X_miss, 1, anyNA),
+   stratify = TRUE)
R> summary(idx)
\end{CodeInput}
\begin{CodeOutput}
========================================================
Train/Test Split Summary
========================================================
Sampling method: Stratified (by missing values)
Total observations: 574
Training set: 459 (80.0%)
Testing set: 115 (20.0%)

--- Missing Values ---
Training set: 108 (23.5%)
Testing set:  27 (23.5%)
Overall:      135 (23.5%)
========================================================
\end{CodeOutput}
\end{CodeChunk}

The training indices are supplied through the \code{idx_train} argument of \code{pr_tree_control()}. For comparison, CART models are fitted using the default settings of \pkg{rpart}, including the complexity parameter \code{cp = 0.01} and the internal 10-fold cross-validation (\code{xval = 10}) used for cost-complexity pruning. To place both methods under comparable tree-growing conditions, we modify only the node-size parameters by setting \code{minsplit = 10} and \code{minbucket = 5}, matching the corresponding stopping criteria adopted by \pkg{PRTree}.

\begin{CodeChunk}
\begin{CodeInput}
R> ctrl <- pr_tree_control(
+   max_terminal_nodes = 30,
+   by_node = TRUE,
+   n_candidates = ncol(X),
+   tiny_sigma = 0,
+   dist = "gamma",
+   shape = 1.5,
+   idx_train = idx$idx_train)
R> ctrlC <- rpart.control(minbucket = 5, minsplit = 10)
\end{CodeInput}
\end{CodeChunk}

The complete and incomplete data sets are then analyzed using both \pkg{PRTree} and CART. For reproducibility, the random seed is fixed before each model fit.

\begin{CodeChunk}
\begin{CodeInput}
R> set.seed(1234)
R> fit <- pr_tree(y = y, X = X, control = ctrl)
R> set.seed(1234)
R> fit_miss <- pr_tree(y = y, X = X_miss, control = ctrl)
R> set.seed(1234)
R> cart <- rpart(y ~ ., data =  data.frame(y, X), control = ctrlC)
R> cp_best <- cart$cptable[which.min(cart$cptable[, "xerror"]), "CP"]
R> fit_cart <- prune(cart, cp = cp_best)
R> set.seed(1234)
R> cart_miss <- rpart(y ~ ., data = data.frame(y, X_miss), control = ctrlC)
R> w_best <- which.min(cart_miss$cptable[, "xerror"])
R> cp_best_miss <- cart_miss$cptable[w_best, "CP"]
R> fit_cart_miss <- prune(cart_miss, cp = cp_best_miss)
\end{CodeInput}
\end{CodeChunk}

To illustrate the effect of the missing-data mechanism, we compare the training root mean squared errors (RMSE) obtained by both methods under complete and incomplete predictor sets.
\begin{CodeChunk}
\begin{CodeInput}
R> rmse <- c(
+   PRTree = sqrt(mean((y - fit$yhat)^2)),
+   PRTree_missing = sqrt(mean((y - fit_miss$yhat)^2)),
+   CART = sqrt(mean((y - predict(fit_cart))^2)),
+   CART_missing = sqrt(mean((y - predict(fit_cart_miss))^2))
+ )
R> round(rmse, 5)
\end{CodeInput}
\begin{CodeOutput}
        PRTree PRTree_missing           CART   CART_missing
       0.62362        1.41710        1.03433        2.73787
\end{CodeOutput}
\end{CodeChunk}
As expected, introducing missing predictor values increases the training error for both methods. However, the increase is considerably smaller for \pkg{PRTree}, reflecting its native probabilistic treatment of missing predictors. In contrast, the default missing-data handling strategy adopted by \pkg{rpart} leads to a substantially larger deterioration in the fitted model for this example.

\begin{figure}[!ht]
\centering
\includegraphics[width=\textwidth]{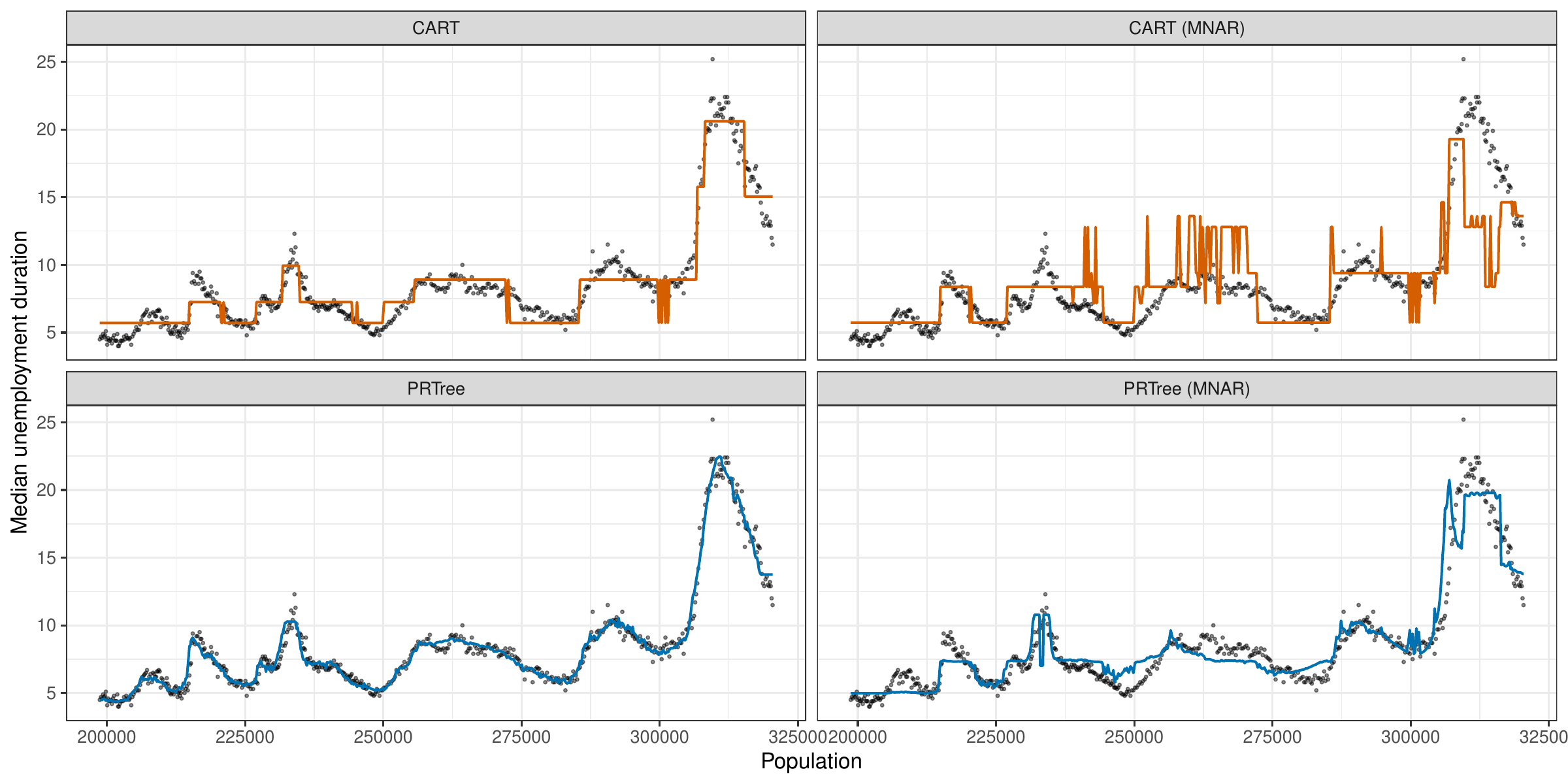}
\caption{\label{fig:missing} Comparison of fitted values obtained by \pkg{PRTree} and CART using complete and incomplete predictor sets. Observations are ordered according to the \code{pop} predictor for visualization purposes. The probabilistic regression trees produce smooth fitted curves, whereas CART yields the characteristic piecewise-constant predictions.}
\end{figure}

To visualize the fitted values, we compare the predictions obtained by \pkg{PRTree} and CART under both complete and incomplete predictor sets. For display purposes, the observations are ordered according to the \code{pop} predictor, while the observed responses are shown as points and the fitted values are represented by colored lines. The resulting plots are displayed in Figure~\ref{fig:missing}.

\begin{CodeChunk}
\begin{CodeInput}
R> library("dplyr")
R> library("ggplot2")

R> ord <- order(X$pop)
R> df_plot <- bind_rows(
+   data.frame(pop = X$pop[ord], observed = y[ord],
+              fitted = fit$yhat[ord], Method = "PRTree"),
+   data.frame(pop = X$pop[ord], observed = y[ord],
+              fitted = fit_miss$yhat[ord], Method = "PRTree (MNAR)"),
+   data.frame(pop = X$pop[ord], observed = y[ord],
+              fitted = predict(fit_cart)[ord], Method = "CART"),
+   data.frame(pop = X$pop[ord], observed = y[ord],
+              fitted = predict(fit_cart_miss)[ord],
+              Method = "CART (MNAR)"))

R> cols <- c("PRTree" = "#0072B2", "PRTree (MNAR)" = "#0072B2",
+           "CART" = "#D55E00", "CART (MNAR)" = "#D55E00")
R> ggplot(df_plot, aes(x = pop)) +
+   geom_point(aes(y = observed),
+              colour = "black", alpha = 0.45, size = 0.9) +
+   geom_line(aes(y = fitted, colour = Method), linewidth = 0.9) +
+   facet_wrap(~Method, ncol = 2) +
+   scale_colour_manual(values = cols) +
+   labs(x = "Population", y = "Median unemployment duration") +
+   theme_bw(base_size = 16) +
+   theme(legend.position = "none")
\end{CodeInput}
\end{CodeChunk}

Although the purpose of this illustration is not to compare predictive performance or establish the superiority of one method over another, it demonstrates that missing predictor values can affect different tree-based methods in different ways. Under the simple MNAR mechanism considered here, the absence of extreme covariate values alters the partitions learned by CART more noticeably than those obtained by \pkg{PRTree}. The probabilistic treatment of incomplete observations enables \pkg{PRTree} to retain information from partially observed covariate vectors, resulting in a smaller change in the fitted values for this example.

\subsection{Cross-validation}

The \pkg{PRTree} package provides a flexible framework for model assessment through repeated data splitting. Since probabilistic regression trees can be configured through different association functions, distribution parameters, missing-data strategies, proxy criteria, and tree-growing options, evaluating multiple model specifications is often desirable. The function \code{pr_tree_cv()} performs cross-validation for a single model specification, using the validation observations to select the smoothing parameter vector (\code{sigma}) and the test observations to estimate predictive performance. When a single candidate smoothing vector is supplied, cross-validation is used only to estimate the test prediction error, whereas setting \code{only_sigma = TRUE} skips the test stage and performs only smoothing-parameter selection. More elaborate model selection procedures can be readily implemented in \proglang{R} by repeatedly calling \code{pr_tree_cv()} under different parameter configurations while reusing the same resampling scheme.

For this example, to ensure a fair comparison across different model specifications, we first generate a common set of stratified folds using \code{train_test_split()}. These folds are subsequently reused by every cross-validation run, so that all candidate models are evaluated under exactly the same resampling scheme.

\begin{CodeChunk}
\begin{CodeInput}
R> set.seed(1234)
R> folds <- train_test_split(
+   n_obs = nrow(X_miss),
+   is_NA = apply(X_miss, 1, anyNA),
+   stratify = TRUE,
+   n_rep = 10
+ )
R> summary(folds)
\end{CodeInput}
\begin{CodeOutput}
========================================================
K-fold Partition Summary
========================================================
Sampling method: Stratified (by missing values)
Total observations: 574
Number of folds: 10

--- Fold Statistics ---
 Fold Size Missing Missing (%)
    1   58      14       24.1%
    2   58      14       24.1%
    3   58      14       24.1%
    4   58      14       24.1%
    5   58      14       24.1%
    6   57      13       22.8%
    7   57      13       22.8%
    8   57      13       22.8%
    9   57      13       22.8%
   10   56      13       23.2%
========================================================
\end{CodeOutput}
\end{CodeChunk}

The cross-validation settings are specified through \code{pr_tree_control_cv()}, which extends \code{pr_tree_control()} with the parameters controlling the resampling procedure. To facilitate comparison with the previous examples, we retain the same tree-growing configuration, including the Gamma association function with shape parameter equal to 1.5, nodewise candidate selection, and a maximum of 30 terminal nodes. The additional arguments define the cross-validation scheme, specifying the previously generated folds and selecting 10-fold cross-validation through \code{method = "kfold"}.
\begin{CodeChunk}
\begin{CodeInput}
R> ctrl_cv <- pr_tree_control_cv(
+   max_terminal_nodes = 30,
+   by_node = TRUE,
+   n_candidates = ncol(X_miss),
+   dist = "gamma",
+   shape = 1.5,
+   fold_idx = folds$fold_idx,
+   method = "kfold"
+ )
R> print(ctrl_cv)
\end{CodeInput}
\begin{CodeOutput}
========================================================
PRTree Control Parameters
========================================================
Sigma grid: not provided (will be generated with size 8)
Multiplier range: [0, 2]
Max terminal nodes: 30
Max depth: 29
Complexity parameter (cp): 0.010
Minimum observations per node: 5
Proportion threshold (prop_x): 0.10
Probability threshold (p_min): 0.050
Fill type: 2
Proxy criterion: both
Number of candidates: 4
By node: TRUE
Distribution: gamma
  with shape = 1.5000
========================================================
Cross-Validation Settings
========================================================
Method: kfold
Number of folds: 10
  Test set: 1 fold per iteration (~10.0% of data).
  Validation (for sigma): 0.20 of the training folds.
Stratification: FALSE
========================================================
\end{CodeOutput}
\end{CodeChunk}

Note that the printed output reports \code{Stratification: FALSE}. This field reflects only the value of the control argument \code{stratify}, whose default is \code{FALSE}. Since the fold assignments were supplied explicitly through \code{fold_idx}, the function does not attempt to infer whether the folds were originally generated using stratified sampling. In this example, however, the supplied folds were obtained with \code{train_test_split(..., stratify = TRUE)}.

To illustrate model comparison, we evaluate the three available missing-data strategies (\code{fill_type = 0}, \code{1}, and \code{2}) while keeping all remaining model parameters fixed.

\begin{CodeChunk}
\begin{CodeInput}
R> cv_fits <- lapply(0:2, function(fill) {
+   set.seed(1234)
+   pr_tree_cv(y = y, X = X_miss, verbose = FALSE,
+     fill_type = fill, control_cv = ctrl_cv
+   )
+ })
R> names(cv_fits) <- paste0("fill_type=", 0:2)
\end{CodeInput}
\end{CodeChunk}

The generic \code{print()} method provides a concise overview of the cross-validation results. It reports the resampling configuration, the average validation and test RMSE across all resampling iterations, and a preview of the selected smoothing parameter vectors (\code{sigma}) together with the corresponding prediction errors from the first iterations.
\begin{CodeChunk}
\begin{CodeInput}
R> print(cv_fits[["fill_type=0"]], digits = 6)
\end{CodeInput}
\begin{CodeOutput}
========================================================
PRTree Cross-Validation Results
========================================================
Method: 10-fold CV
Test set: 1 fold per iteration (~10.0% of data).
Validation (for sigma): 0.20 of the training data/folds.
Features: 4

--- Sigma Selection ---
Validation RMSE (mean +/- sd): 1.508746 +/- 0.304146

--- Test Error ---
Test RMSE (mean +/- sd): 1.871837 +/- 0.905932

First 5 iterations/folds (sigma values and RMSE):
 iter     pce     pop psavert unemploy RMSE_val RMSE_test
    1 1504.47 16423.2 1.32176  926.231  1.05381   2.04032
    2 3675.77 40363.9 3.24292 2327.238  1.25056   1.84173
    3 1501.90 16481.8 1.30611  946.406  1.89663   1.59093
    4 5326.23 57815.4 4.57365 3253.802  1.69703   1.35986
    5 5301.10 57932.5 4.58551 3315.714  1.60411   1.15576
========================================================
\end{CodeOutput}
\end{CodeChunk}

More detailed information is obtained through the generic \code{summary()} method. Besides reporting the cross-validation configuration, the summary provides descriptive statistics of the validation and test RMSE across all resampling iterations, as well as summary statistics of the selected smoothing parameter vectors (\code{sigma}) for each predictor. Here, the summary object is explicitly passed to \code{print()} so that the number of displayed significant digits can be controlled.
\begin{CodeChunk}
\begin{CodeInput}
R> print(summary(cv_fits[["fill_type=0"]]), digits = 5)
\end{CodeInput}
\begin{CodeOutput}
========================================================
Summary of PRTree Cross-Validation
========================================================
Cross-Validation Configuration:
  Method: 10-fold CV
  Test set: 1 fold per iteration (~10.0% of data).
  Validation (for sigma): 0.20 of the training folds.
  Features: 4
  Stratification: FALSE

--- RMSE Summary ---
Validation Set (used for sigma selection):
    Min     Q1 Median   Mean     Q3    Max      SD  N
 1.0538 1.2893 1.5544 1.5087 1.6875 1.9146 0.30415 10

Test Set (unseen data):
    Min     Q1 Median   Mean     Q3    Max      SD  N
 1.1558 1.3729 1.6864 1.8718 1.9687 4.2918 0.90593 10

--- Selected Sigma Summary (by feature) ---
               Min       Mean    Median        Max         SD
pce       744.6079  3079.4882  2960.047  5326.2296  1655.0720
pop      8171.2657 33651.7516 32483.339 57932.5163 18032.1115
psavert     0.6437     2.6772     2.604     4.5855     1.4283
unemploy  463.1093  1922.3504  1873.528  3315.7136  1027.9653
========================================================
\end{CodeOutput}
\end{CodeChunk}

\begin{figure}[!ht]
\centering
\includegraphics[width=\textwidth]{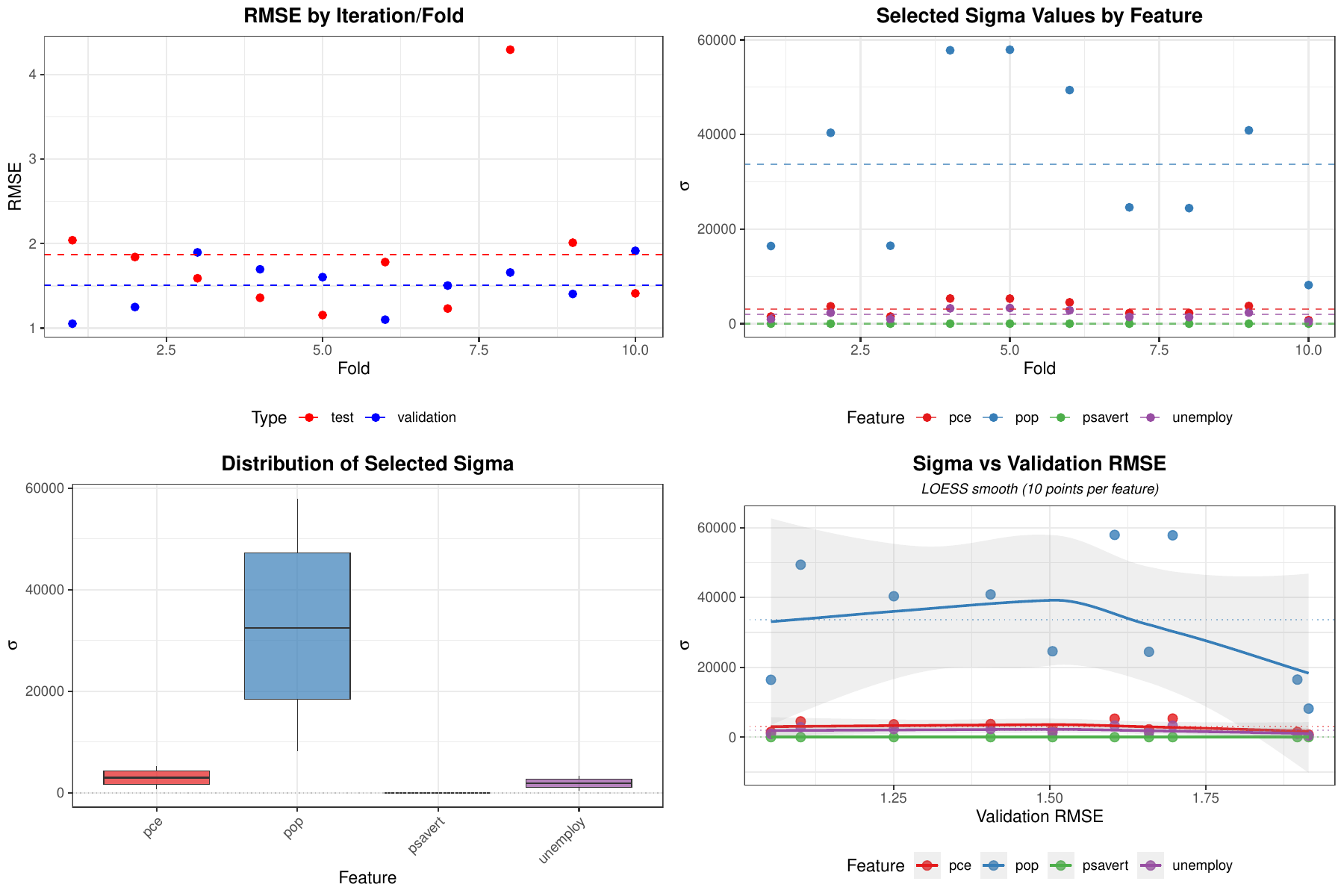}
\caption{\label{fig:cv_plot} Diagnostic plots produced by \code{plot()} for a cross-validation object. The panels display the validation and test RMSE across cross-validation iterations (upper left), the selected smoothing parameter vectors for each predictor (upper right), the empirical distribution of the selected smoothing parameters (lower left), and the relationship between the selected smoothing parameters and the corresponding validation RMSE (lower right).}
\end{figure}

The generic \code{plot()} method provides several graphical summaries of the cross-validation results, including the validation and test RMSE across resampling iterations, the selected smoothing parameter vectors, their empirical distributions, and the relationship between the selected smoothing parameters and the corresponding validation RMSE. The resulting plots are shown in Figure~\ref{fig:cv_plot}. Individual panels can be displayed by specifying the corresponding indices through the \code{which} argument. This is particularly useful when the selected smoothing parameters have markedly different magnitudes across predictors, as illustrated in Figure~\ref{fig:cv_plot}, where the last three panels are dominated by the predictor with the largest smoothing parameter.
\begin{CodeChunk}
\begin{CodeInput}
R> plot(cv_fits[["fill_type=0"]])
\end{CodeInput}
\end{CodeChunk}

Since the cross-validation results are returned as ordinary \proglang{R} objects, they can be readily manipulated using standard \proglang{R} tools. As a simple illustration, the following code constructs a graphical comparison of the predictive performance of the three missing-data strategies by combining the validation and test RMSE obtained across all cross-validation repetitions.
\begin{CodeChunk}
\begin{CodeInput}
R> df <- do.call(rbind, lapply(seq_along(cv_fits), function(i) {
+  fit <- cv_fits[[i]]
+  fill <- as.character(i - 1)
+  rbind(data.frame(RMSE = fit$rmse_by_rep$test, Fill = fill, type = "test"),
+        data.frame(RMSE = fit$rmse_by_rep$validation, Fill = fill,
+                   type = "validation"))
}))
R> ggplot(df, aes(x = Fill, y = RMSE, fill = type)) +
+    geom_boxplot(position = position_dodge(width = 0.8), width = 0.6) +
+    labs(x = "Fill", y = "RMSE", fill = "Type") +
+    theme_bw()
\end{CodeInput}
\end{CodeChunk}
\begin{figure}[!ht]
\centering
\includegraphics[width=.7\textwidth]{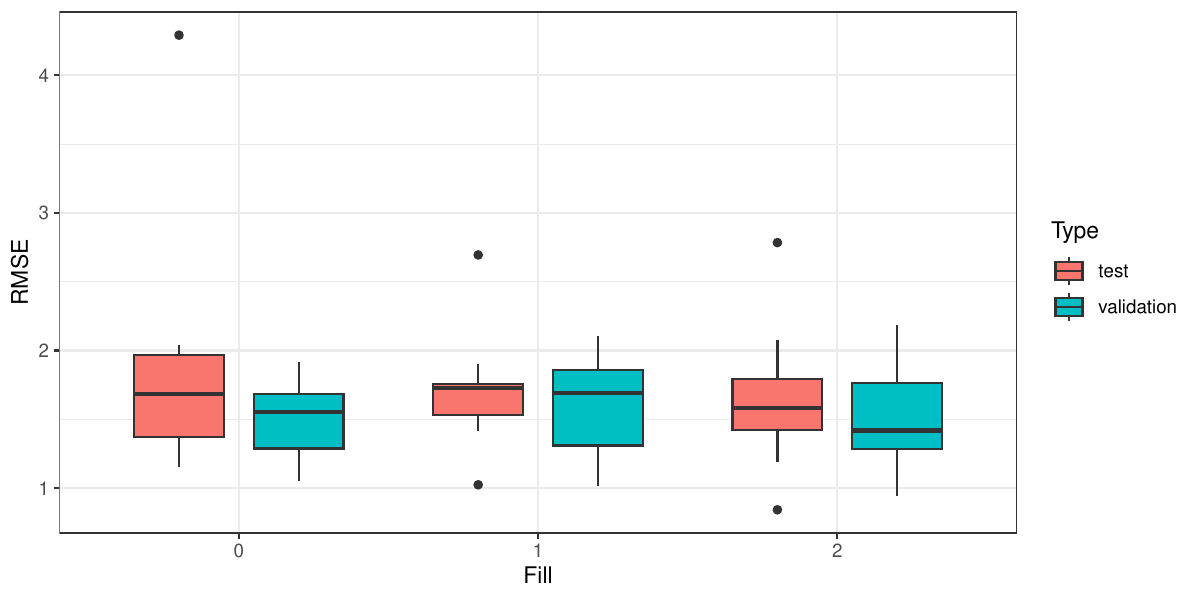}
\caption{\label{fig:cv_filltype} Graphical comparison of the three missing-data strategies based on the validation and test RMSE obtained over the 10 cross-validation repetitions. The boxplots summarize the empirical distribution of the prediction errors for each \code{fill\_type}.}
\end{figure}

Although this illustration compares only the three missing-data strategies, the same workflow can be used to compare any collection of model specifications, including different association functions, distribution parameters, tree-growing options, and proxy criteria. By reusing the same resampling partitions across all candidate models, differences in predictive performance can be attributed to the model specification rather than to the randomness of the resampling procedure.

\FloatBarrier

\section{Summary and discussion} \label{sec:summary}

This paper introduced \pkg{PRTree}, an \proglang{R} package for fitting probabilistic regression trees. The package provides a comprehensive implementation of the methodology, including model fitting, prediction, visualization, cross-validation, and support for incomplete predictor data through multiple missing-data strategies. The implementation follows the standard \proglang{R} object-oriented framework, offering familiar interfaces through generic methods such as \code{print()}, \code{summary()}, \code{plot()}, and \code{predict()}.

Beyond the core fitting procedure, the package includes tools for constructing reproducible training and validation partitions, performing repeated cross-validation, selecting smoothing parameters, and comparing alternative model specifications under a common resampling scheme. These utilities facilitate both methodological research and practical applications of probabilistic regression trees.

Future developments of the package will focus on extending the methodology to additional splitting criteria, alternative probabilistic association functions, and specialized visualization and diagnostic tools. We also expect the package to serve as a platform for the implementation and dissemination of new methodological developments in probabilistic tree-based regression.

\section*{Computational details}

The results presented in this paper were obtained using \proglang{R}~4.5.1 with the packages \pkg{PRTree}, \pkg{ggplot2}, and \pkg{rpart}. The \pkg{PRTree} package relies on compiled FORTRAN routines for the computationally intensive steps of tree construction, while \pkg{ggplot2} was used to produce the figures shown throughout the paper. \proglang{R} and all packages used are available from the Comprehensive \proglang{R} Archive Network (CRAN) at \url{https://CRAN.R-project.org/}.

\section*{Acknowledgments}

T.S. Prass and G. Pumi gratefully acknowledge the financial support received by the Conselho Nacional de Desenvolvimento Cient\'ifico e Tecnol\'ogico -- CNPq Brasil  -- Bolsa de Produtividade em Pesquisa - Proc. 305886/2025-8 (Prass) and 303281/2025-1 (Pumi). A.S. Neimaier gratefully acknowledge the financial support received by the Coordena\c{c}\~ao de Aperfei\c{c}oamento de Pessoal de N\'ivel Superior – Brasil (CAPES) – Finance Code 001.

\bibliographystyle{jss}
\bibliography{refs}

@inproceedings{alkhoury2020,
	title        = {Smooth And Consistent Probabilistic Regression Trees},
	author       = {Sami Alkhoury and Emilie Devijver and Marianne Clausel and Myriam Tami and Eric Gaussier and Georges Oppenheim},
	year         = 2020,
	booktitle    = {Advances in Neural Information Processing Systems},
	publisher    = {Curran Associates, Inc.},
	volume       = 33,
	pages        = {11345--11355},
	editor       = {H. Larochelle and M. Ranzato and R. Hadsell and M.F. Balcan and H. Lin},
	url          = {https://dl.acm.org/doi/pdf/10.5555/3495724.3496676}
}

@Book{breiman84,
  Title                    = {Classification and Regression Trees},
  Author                   = {Leo Breiman and Jerome Friedman and Charles J. Stone and R.A. Olshen},
  Publisher                = {Chapman and Hall/CRC},
  Year                     = {1984}
}

@inproceedings{irsoy2012,
	title        = {Soft decision trees},
	author       = {İrsoy, Ozan and Yıldız, Olcay Taner and Alpaydın, Ethem},
	year         = 2012,
	month        = nov,
	booktitle    = {Proceedings of the 21st International Conference on Pattern Recognition (ICPR2012)},
	pages        = {1819--1822},
	issn         = {1051-4651}
}

@article{medeiros2008,
	title        = {Tree-structured smooth transition regression models},
	author       = {Joel Corrêa {da Rosa} and Alvaro Veiga and Marcelo C. Medeiros},
	journal      = {Computational Statistics \& Data Analysis},
	volume       = {52},
	number       = {5},
	pages        = {2469--2488},
	year         = 2008,
	ISSN         = {0167-9473}
}

@misc{PRTree2026_theory,
	title        = {Handling Missing Data in Probabilistic Regression Trees},
	author       = {Taiane Schaedler Prass and Alisson Silva Neimaier and Guilherme Pumi},
	year         = 2026,
    url          = {https://doi.org/10.48550/arXiv.2608.06195}    
}

@manual{rpart,
	title        = {rpart: Recursive Partitioning and Regression Trees},
	author       = {Terry Therneau and Beth Atkinson},
	year         = 2019,
	url          = {https://CRAN.R-project.org/package=rpart},
	note         = {{R package version 4.1-15}}
}

\end{document}